\documentclass{aa}

\usepackage[varg]{txfonts}
\usepackage[font=footnotesize,labelfont={bf,sc,small},figurename=Fig.]{caption}
\usepackage[hidelinks]{hyperref}
\graphicspath{{./figures/}}
\usepackage{subfig}
\usepackage{rotate}
\usepackage{appendix}
\usepackage{xcolor}
\usepackage{varioref}
\usepackage{ulem}
\usepackage{natbib}
\bibpunct{(}{)}{;}{a}{}{,} 
\hypersetup{draft}
\newcommand{\mws}[0]{\mbox{S$_{\rm MWO}$}}
\newcommand{\scairt}[0]{\mbox{S$_{\rm Ca~IRT}$}}
\newcommand{\epsEri}{$\epsilon$\,Eri~}
\usepackage{multirow}
\usepackage{amstext}

\begin{document}

\title{A multi-wavelength view of the multiple activity cycles of $\epsilon$~Eridani}
\author{B. Fuhrmeister\inst{\ref{inst1}}, M. Coffaro\inst{\ref{inst2}}, B. Stelzer\inst{\ref{inst2},\ref{inst3}}, M. Mittag\inst{\ref{inst1}} 
\and S.~Czesla\inst{\ref{inst1}}
\and P.~C. Schneider\inst{\ref{inst1}}
}


\institute{Hamburger Sternwarte, Universit\"at Hamburg, Gojenbergsweg 112, D-21029 Hamburg, Germany\\ 
\email{bfuhrmeister@hs.uni-hamburg.de}\label{inst1}
\and
Institut f\"ur Astronomie und Astrophysik T\"ubingen, 
Eberhard Karls Universit\"at T\"ubingen,  Sand 1, D-72076 T\"ubingen, Germany \label{inst2} 
\and 
INAF - Osservatorio Astronomico di Palermo, Piazza del Parlamento 1, I-90134 Palermo, Italy\label{inst3}
}

\date{Received / Accepted}

\abstract{$\epsilon$ Eridani is a highly active young K2 star with an activity cycle of about
  three years established using \ion{Ca}{ii} H\& K line index measurements (\mws). This
  relatively short cycle has been demonstrated to be consistent with  X-ray and magnetic
  flux measurements. 
  Recent work suggested a change in the cyclic behaviour. Here we report
new X-ray flux and \mws\ measurements and also include   \mws\ measurements from the  historical Mount Wilson program. This results in an observational time baseline of over 50 years
  for the \mws\ data and of over 7 years in  X-rays. Moreover, we include \ion{Ca}{ii}
  infrared triplet (IRT) index measurements (\scairt) from 2013-2022 in our study. 
  With the extended X-ray data set, we can now detect the short cycle for the first time using a 
  periodogram analysis. Near-simultaneous  \mws\ data and  X-ray fluxes, which are offset by 20 days at most, 
  are moderately strongly correlated when only the lowest activity state (concerning short-term
  variability) is considered in both diagnostics.
   In the \mws\ data, we find strong evidence for a much longer cycle of about 34 years and an 11-year cycle instead of the formerly proposed 
   $12$-year cycle in addition
  to the known 3-year cycle. The superposition of the three periods
  naturally explains the recent drop in \mws\ measurements. The two shorter cycles 
  are also detected in
 the \scairt\ data, although the activity cycles exhibit lower amplitudes in the
  \scairt\ than in the \mws\ data. 
   Finally, the rotation period of \epsEri can be found more frequently in the \mws\ as well as in the
  \scairt\ data  for times near the minimum of the long cycle. This may be explained by a scenario in which the 
  filling factor for magnetically active regions near cycle maximum is too high to allow for notable short-term variations.}

\titlerunning{Changing activity cycle of $\epsilon$ Eridani}
\authorrunning{B. Fuhrmeister et~al.}
\maketitle

\section{Introduction}\label{sec:intro}

$\epsilon$~Eridani ($\epsilon$~Eri; HD~22049) is a K2\,V star at
a distance of 3.2~pc \citep{BailerJones2021}.  
It is known to host one planet with a semi-major axis of 3.4 AU and a planetary candidate at 40 AU \citep{Hatzes2000, Quillen2002}. The latter was inferred by its imprint on the morphology
of the dust ring around $\epsilon$~Eri.
The stellar age was determined by \cite{Barnes2007} to be 440~Myr from gyrochronology, while \citet{Marsden2015} found an age of 480 Myr from 
its chromospheric activity level.
The chromospheric activity of $\epsilon$\,Eri has been studied extensively, starting with measurements from the
the  Mount Wilson S-index (\mws) program in 1968 \citep{Wilson1978}, which
provided
a 
purely instrumental index covering the emission cores of the \ion{Ca}{ii} H\&K
lines. 

The
Mount Wilson HK project discovered activity cycles for many F- to K-type stars
\citep{Baliunas1995} and also determined a rotation period of 11.10$\pm$0.03 days and
an activity cycle of approximately five years for $\epsilon$~Eri \citep{Gray1995}.
Later, \citet{Froehlich2007} revealed differential rotation with two rotation
periods measured  at $P_{\rm rot}$=11.35~days and $P_{\rm rot}$=11.555~days. Moreover,
\citet{Metcalfe2013} reanalysed the Mount Wilson data and also included data from the Small and Moderate Aperture Research
Telescope System (SMARTS) southern  HK project. The authors proposed two cycles in $\epsilon$~Eri, one with
2.95$\pm$0.03\,yr, the other with 12.7$\pm$0.3\,yr length. 
With these properties, $\epsilon$~Eri is one of the youngest stars with a known activity cycle. \citet{Metcalfe2013}
also found 
 peaks between 3 and 7 years in their periodogram analysis, which they discarded, and a
peak at 20-35 years, which they
excluded because 
this time span is similar to the length of the adopted data set.

The 3-year cycle 
was also studied by other means. For example, \citet{Scalia2018} found that the integrated longitudinal magnetic field shows a period comparable
to the 3-year \mws\ cycle. In a more detailed analysis, \citet{Jeffers2022} found that
the net axis-symmetric component of the toroidal magnetic field correlates with the
12-year calcium cycle modulated by the short \mws\ cycle.
Moreover, $\epsilon$~Eri is one of the few stars for which a cycle was detected in X-rays.
Using {\it XMM-Newton} observations, \cite{Coffaro2020} found the X-ray variations
to be consistent with the 3-year \mws\ cycle. 
 Based on modelling of $\epsilon$\,Eri  X-ray spectra with observations of our Sun, treating it in the framework of {the-Sun-as-an-Xray-star} (see references in \cite{Coffaro2020}), they found
that the X-ray
cycle is characterised by changes in the filling factor of magnetically active structures ranging between 60\%
to 90\%  
from minimum to maximum activity level. This high coronal
filling factor throughout the whole activity cycle also explains the low amplitude of the X-ray cycle
compared to that of the Sun. X-ray cycle maxima also exhibit high filling factors of flaring regions. This finding, indirectly inferred from comparison with 
 solar data treated with the Sun-as-an-Xray-star technique,
 is consistent
with the fact that resolved flares in the 
X-ray light curves are predominantly found 
near
cycle maxima. Nevertheless, \citet{Burton2021} found a flare at millimeter wavelengths
in data taken by the ALMA instrument at the beginning of 2015. In this time, the activity of
$\epsilon$~Eri was in a minimum state.

\cite{Coffaro2020} also  noted that \mws\ data collected with the TIGRE
telescope \citep{Schmitt2014} seems to indicate a change in the cycle behaviour in observations   
since 2018,
as the expected 2019 maximum was not seen in the \mws\ data, but only in
the X-ray measurements. The component of the magnetic field  measured by 
\citet{Jeffers2022} also indicates lower values in 2019, which \citet{Jeffers2022}
interpreted as being caused by the superposition of the magnetic field components of the 3- and 12-year cycles.

  Here we extend the time-series of the \mws\ data  from the work by \citet{Coffaro2020} both to older (using the Mount Wilson program data) and to newer data
up to February 2022 (from the TIGRE telescope).
This collection of more than 50 years of \mws\ observations, corresponding to dozens of
the $3$-year cycles,
allows us to search for multiple and especially very long cycles for 
$\epsilon$\,Eri, such as those that were 
presented for other stars, for example, by \citet{Brandenburg2017}. 
It also enables us to study cycle length variations, which are well
known for the Sun \citep{Hathaway2015, Ivanov2021}, in another star. Dynamo theory can explain cyclic activity behaviour as such, but different shapes or lengths of individual cycles are not well understood or even constrained yet 
because it is difficult to monitor the stars for a sufficiently long time.

We intend to study the ongoing transition in chromospheric activity by adding
more data to the \mws\ measurements 
as well as activity indices
defined for the \ion{Ca}{ii} infrared triplet (IRT) lines  measured with
the TIGRE telescope.
We also want to extend the studies of coronal
activity by adding the latest X-ray data observed with \textit{XMM-Newton}.

\section{Observations and data analysis}
$\epsilon$~Eri has been monitored  with different instruments in optical spectroscopy with a
total time baseline of more than $50$\,years and for 7 years in the X-ray waveband. 

In the following, we present the \ion{Ca}{ii} $S_{\rm MWO}$ data acquired by different
ground-based observatories (Sect. \ref{sec:cadata}) and the X-ray monitoring campaign
(Sect. \ref{sec:xrayobs}) with the \textit{XMM-Newton} satellite.
For the X-ray data, we also include an analysis of the short-term
variations of the coronal flux due to flares. These variations potentially contaminate long-term variations and should therefore
be excluded from a determination of the activity cycle.

\subsection{Optical data}
\label{sec:cadata}
The $S_{\rm MWO}$ values used here originate from different observatories; the
oldest are from the Mount Wilson program itself\footnote{The data are accessible
through \tt{https://dataverse.harvard.edu/dataset.xhtml}}. 
Further optical spectroscopic observations for $\epsilon$~Eri were obtained within the California Planet Search
(CPS) program at the Keck and Lick Observatories \citep{Isaacson2010}, with the Solar-Stellar Spectrograph
located at the Lowell Observatory, and with the SMARTS instrument at the Cerro Tololo 
Interamerican Observatory. From these data, $S_{\rm MWO}$ values were derived and already
published in \citet{Coffaro2020}, where a more detailed description of these
data can also be found. \citet{Coffaro2020} also presented $S_{\rm MWO}$ values from the TIGRE
telescope located in Guanajato, Mexico \citep{Schmitt2014, Gonzalez2022}. The optical monitoring
with the TIGRE telescope of $\epsilon$~Eri is still ongoing, and we present here
data until mid February 2022. Moreover, we include here activity indices $S_{\rm Ca IRT}$ that were calculated from
the spectra of the TIGRE telesope for the
\ion{Ca}{ii} IRT lines as described by \citet{Mittag2017}.
These lines are located at 8498, 8542, and 8662 \AA,\, and the respective line and continuum wavebands used for the calculation of $S_{\rm Ca IRT}$ can be found in \citet{Mittag2017}. These lines can be used for
chromospheric activity studies, as was shown for example by their correlation with \ion{Ca}{ii}\,H\&K in flux-flux relations for F to M dwarfs studied by
\citet{Martinez-Arnaiz2011}.

Timing information of all 
Ca\,{\sc II} data can be found in Table~\ref{Tab:caobs}.
Since especially the Mount Wilson program included multiple observations per night
(usually three; but single nights have $>$200 observations), we computed the mean of
these observations. We therefore have one (mean) $S_{\rm MWO}$ measurement per night,
which leaves us with 675 Mount Wilson program measurements and  a total of 1578
measurements taken between 1967 and 2022. From these, we clipped three apparent
outliers with one \mws $< 0.36$ and two measurements $>0.63$. One of the two high \mws\ values is from the CPS program, and the other is an individual measurement from the Mount Wilson program.  
Both are likely
caused by flaring activity. Since the data from the Mount Wilson
and SMARTS programs do not have errors assigned and because part of the Lowell observatory data
have exceptionally low errors, we used an error of 3\%\ for all data because this is the median error of the TIGRE data.
Of this whole data set, we considered the two subsets separately, namely the
Mount Wilson data and the TIGRE data. These are separated in time by about 20 years.
When we considered only the TIGRE data, the errors obtained from the pipeline
were used.

We excluded $S_{\rm Ca IRT}$  
data from between 30 November 2014 and 16 May 2015
because a different camera was used for the red spectrograph arm of TIGRE during 
that period.

\begin{table}
        \caption{\label{Tab:caobs} Basic information about the \ion{Ca}{ii} observations.  }
\footnotesize
\begin{tabular}[h!]{lccccc}
\hline
\hline
\noalign{\smallskip}

Telescope           & No.   & JD first & JD last &covered \\
 & spectra & [day] & [day] & years\\
\noalign{\smallskip}
\hline
\noalign{\smallskip}
Mount Wilson & 4336 & 2439786.8 & 2449771.6&1967-1995\\
CPS &   168 & 2452267.7 & 2455415.1&2001-2010\\
SMARTS &  146 & 2454334.7 & 2456324.0&2007-2013 \\
Lowell Obs. &  267 & 2449258.9 & 2458154.5&1993-2018\\
TIGRE &  322 & 2456518.9 & 2459624.6&2013-2022\\
all data & \\
1 day binning & 1578 \\
\hline\noalign{\smallskip}
TIGRE \ion{Ca}{ii} IRT & 303 & 2456518.9 & 2459624.6\\
excluded times & & 2456992.0 & 2457159.0 \\
\noalign{\smallskip}
\hline

\end{tabular}
\normalsize
\end{table}

\subsection{X-ray observations}
\label{sec:xrayobs}
At X-ray wavelengths, a monitoring campaign started in August 2015 (PI: B.Stelzer) 
using the \textit{XMM-Newton} satellite.
It consists of snapshots with durations between 7.6 and 21.5 \,ks, repeated roughly
every $\text{six}$ months.
Prior to this campaign,  
\epsEri had been observed twice with {\it XMM-Newton},
in January 2003 and February 2015 (PIs: B. Aschenbach and K. France). 
\textit{XMM-Newton} observed $\epsilon$~Eri employing all X-ray 
telescopes on board. Hence, we have EPIC (pn+MOS) and RGS data 
products. 
For this work, we made use of EPIC/pn data. EPIC/MOS provides a lower
signal-to-noise ratio and does not yield additional information for our study.
The analysis of the RGS data is deferred to a future work. As $\epsilon$~Eri
is a bright star ($m_V \sim 3$), the Optical Monitor on board of
\textit{XMM-Newton} cannot be used. 

The first three years of monitoring (2015-2018), plus the 2003 and 2015 archival data, 
in total nine observations, were presented and analysed by \citet{Coffaro2020}.
These authors discovered 
the X-ray cycle of $\epsilon$\,Eri.
After 2018, the X-ray monitoring continued until January 2022, providing seven
new  observations. We follow the approach described by \citet{Coffaro2020} and justified above
that we extracted data from the EPIC/pn detector alone.  
The observing log of all  available {\it XMM-Newton} 
pointing observations
is given in Table~\ref{tab:obslog}. For later reference throughout this article, we define in col.~1 a running number for  each observation following chronological order.

\begin{table}
        \caption{\label{tab:obslog} X-ray observations with \textit{XMM-Newton}. }
\footnotesize
\begin{tabular}[h!]{ccccc}
\hline
\hline
\noalign{\smallskip}
    Obs. &Date & Rev. & Science Mode & Exposure time \\
    no.  &      &  & (EPIC/pn)   &   [ksec]       \\
    \noalign{\smallskip}
    \hline    
    \noalign{\smallskip}
    1& 2003-01-19 & 0570 & Full window &13.4 \\
    2&      2015-02-02 & 2775& Large window&20.0 \\
    3&      2015-07-19 & 2858& Small window&8.0 \\
    4&      2016-02-01 & 2957& Small window&9.3 \\
    5&      2016-07-19 & 3042& Small window&11.0\\
    6&      2017-01-16 & 3133& Small window&7.6 \\
    7&      2017-08-26 & 3244& Small window&10.3\\
    8&      2018-01-16 & 3316& Small window&8.0 \\
    9&      2018-07-20 & 3408& Small window&21.5\\
   10&      2019-01-19 & 3500& Small window&18.4\\
   11&      2019-07-19 & 3591& Small window&8.8\\
   12&      2020-01-19 & 3683& Small window&7.9\\
   13&      2020-07-22 & 3776& Small window&9.9\\
   14&      2021-01-17 & 3866& Small window&8.9\\
   15&      2021-08-07 & 3967& Small window&8.0\\
   16&      2022-01-17 & 4049& Small window&13.9\\
\noalign{\smallskip}
\hline

\end{tabular}
\normalsize
\end{table}

\subsection{Analysis of XMM-Newton data}

The focus of this work is to study the long-term variability of $\epsilon$\,Eri.
  However, this requires an assessment of short-term  variations that could
  modify the time-averaged flux of each snapshot observation.
  In order to obtain reliable flux measurements for comparison to the optical data,
we identified flaring episodes and excluded them from the further data analysis. This process
is described in the following.  

\subsubsection{Short-term variability in EPIC/pn light curves}

EPIC/pn data were extracted with the software called\texttt{} 
Science Analysis System (SAS; version 17.0.0). The standard \texttt{SAS} tools were
applied to filter event lists of each observation and produce the images, and also to extract the light curve and spectrum  of $\epsilon$\,Eri for each individual observation. 
Then we identified flaring states using a timing
analysis and the hardening of the X-ray spectrum, which is caused by higher
temperatures during flares.

First, 
the EPIC/pn light curves of $\epsilon$~Eri were extracted in the 
$0.2-2.0$~keV energy band, and they were binned with a bin size of $300$\,s. 
Following the approach of \citet{Coffaro2020}, we searched for short-term variability 
in each light curve with the software \texttt{R} and its library \texttt{changepoint} \citep{changepoint}. 
 Through this analysis, \citet{Coffaro2020} found that four out of nine observations  to display short-term 
variability. For the seven new observations, short-term variability was 
identified in five light curves. In Fig.~\ref{fig:hr_lc} we show all variable light curves. The segments are identified by \texttt{changepoint} and are drawn as dash-dotted black lines. For comparison, we also show the light curves with a constant count rate in Fig. \ref{fig:hr_lc_w}. They represent the lowest count rates, that is, the most quiescent observations of \epsEri.

This short-term variability might be related to flares. To search for evidence
  of heating episodes, we generated light curves of the hardness ratio (HR) for observations that the software \texttt{R} flagged as variable.
The HR is calculated as 
\begin{equation}
    HR = \dfrac{C_h - C_s}{C_h + C_s}
    \label{eq:hr}
,\end{equation}
where $C_h$ and $C_s$ are the count rates in a hard and a soft energy band, respectively.
We chose the 
energy range $0.2-1.0$~keV as soft band and $1.0-2.0$~keV as hard band.
The HR light curves of each observation are shown in the bottom panels of each plot 
in Fig.~\ref{fig:hr_lc}. In each of these HR light curves, we looked for variability with the software \texttt{R,} analogously to the treatment of the light curves outlined above. The corresponding segments
are indicated with the dash-dotted green lines in each bottom panel.


\subsubsection{Spectral analysis of EPIC/pn data}\label{subsubsect:analysis_epic_spectrum}

We analysed the EPIC/pn spectra of each observation with the software \texttt{xspec}. We chose a 3-T APEC model with global metal 
abundances frozen at $0.3 \ Z_\odot$\footnote{\citet{Coffaro2020} found from the analysis of observations from 2015 to 2018 that when $Z$ is free to vary during the fitting procedure, it spans the range $0.2-0.4 \ Z_\odot$.
In that paper, we therefore decided to keep $Z$ frozen to an average value of $0.3\,Z_\odot$. Here, we followed the same approach. The accurate determination of the abundances of individual elements on the basis of the RGS spectra will be discussed elsewhere.}.
 We did not include photoelectric absorption because $\epsilon$~Eri is 
 very nearby,
  thus 
 no interstellar absorption is expected.
 The best-fitting model provides three emission measures (EM) and three
   thermal energy components (kT), and the results are given in
   Table~\ref{tab:epicpn_spectralparams} for all observations presented here for the
   first time, while an analogous table is found in \cite{Coffaro2020} for the
   data up to 2018.

\begin{table*}
  \tiny
\caption{\label{tab:epicpn_spectralparams} Best-fit spectral parameters of  
all \textit{XMM-Newton} EPIC/pn observation of \epsEri\, that are presented here for the first time}.
\begin{tabular}{cccccccccccc}
  \hline
  \hline
\noalign{\smallskip}
Obs  &  $kT_1$          & $kT_2$          & $kT_3$          & $logEM_1$        & $logEM_2$        & $logEM_3$        & Flux                            & $L_{\rm X}$ & $T_{\rm av}$ & $\overline{\chi}^2$  \\
No. &                   &                 &                 &                  &                 &                   & (0.2-2\,keV)                  &  (0.2-2\,keV)  & &  \\
    & [keV]         & [keV]         & [keV]         & [cm$^{-3}$]        & [cm$^{-3}$]        & [cm$^{-3}$]        & [10$^{-11}$ erg\,cm$^{-2}$\,s$^{-1}$] & [10$^{28}$ erg\,s$^{-1}$] & [keV]  &\\ 
\hline
\noalign{\smallskip}
                10 & 0.12 $\pm$ 0.01& 0.31 $\pm$0.01 & 0.71 $\pm$0.02 & 50.70 $\pm$0.09& 50.93 $\pm$0.02& 50.34 $\pm$0.06 & 1.25 $\pm$0.02 & 1.54$\pm$ 0.01 &0.31$\pm$ 0.02& 1.39\\
        11 & 0.21 $\pm$ 0.04& 0.43 $\pm$0.07 & 0.87 $\pm$0.07 & 50.87 $\pm$0.05& 50.80 $\pm$0.07& 50.81 $\pm$0.04 & 1.85 $\pm$0.03 & 2.27$\pm$ 0.04 &0.49$\pm$ 0.03& 1.01\\
        12 & 0.14 $\pm$ 0.02& 0.32 $\pm$0.02 & 0.68 $\pm$0.05 & 50.63 $\pm$0.11& 50.97 $\pm$0.05& 50.34 $\pm$0.08 & 1.28 $\pm$0.03 & 1.57$\pm$ 0.04 &0.32$\pm$ 0.02& 0.95\\
        13 & 0.17 $\pm$ 0.03& 0.33 $\pm$0.02 & 0.72 $\pm$0.03 & 50.68 $\pm$0.15& 50.93 $\pm$0.12& 50.57 $\pm$0.06 & 1.50 $\pm$0.03 & 1.84$\pm$ 0.04 &0.37$\pm$ 0.02& 0.98\\
        14 & 0.17 $\pm$ 0.03& 0.34 $\pm$0.02 & 0.75 $\pm$0.02 & 50.71 $\pm$0.07& 51.02 $\pm$0.07& 50.68 $\pm$0.05 & 1.80 $\pm$0.03 & 2.21$\pm$ 0.04 &0.40$\pm$ 0.02& 1.26\\
        15 & 0.14 $\pm$ 0.02& 0.32 $\pm$0.02 & 0.71 $\pm$0.04 & 50.67 $\pm$0.10& 50.94 $\pm$0.08& 50.37 $\pm$0.08 & 1.30 $\pm$0.03 & 1.59$\pm$ 0.04 &0.33$\pm$ 0.02& 0.90\\
        16 & 0.15 $\pm$ 0.02 & 0.31 $\pm$ 0.04 &0.79 $\pm$ 0.02 &50.62 $\pm$ 0.08 & 50.91 $\pm$ 0.12 & 50.52 $\pm$ 0.05 &1.34 $\pm$0.02 & 1.65$\pm$ 0.03 & 0.37$\pm$ 0.02 & 1.26\\

        \hline
\end{tabular}
\normalsize
\end{table*}

 We calculated the EM weighted average thermal energy as
\begin{equation}
    kT_{\rm av} = \dfrac{\sum _i kT_i \cdot EM_i}{\sum _i EM_i}
,\end{equation}
where
$i=1,2,3$ is the index for each spectral component.
We also calculated the X-ray fluxes observed at Earth ($F_{\rm x}$) in the soft energy band $0.2-2.0$~keV,
and the X-ray luminosity $L_{\rm X}$.
In Table~2 we report the $kT_{\rm av}$ and $F_{\rm X}$ values obtained from the spectral fitting.
        The errors given there are the statistical errors from the fitting process, which
        represent the 95\% confidence level. 
The variability in the X-ray light curves of the individual observations indicates 
larger flux changes, however. Following \cite{SanzForcada2019}, we therefore used the standard deviation of the
flux variations within a given light curve as the flux error for the timing analysis.

For each of the observations that were flagged as variable by the software \texttt{R,} the same
spectral analysis as described above for the time-average of each observation was
repeated for two time intervals.  That is, we extracted two
separate spectra, one spectrum referring to the quiescent state of the observation
(the segment with the lowest count rate in the light curve), and another within the flare-like event (the segment identified by \texttt{R} with the highest count rate). 
In Table~\ref{tab:QF_v} we report the $kT_{\rm av}$ and $F_{\rm X}$ values of the best
fit to these different activity states. An increase in the average 
plasma
thermal energy is found for all flare-like time intervals, but in most cases, it is
only marginally significant. 

\begin{figure}[!htbp]
    \centering
    \includegraphics[width=0.5\textwidth]{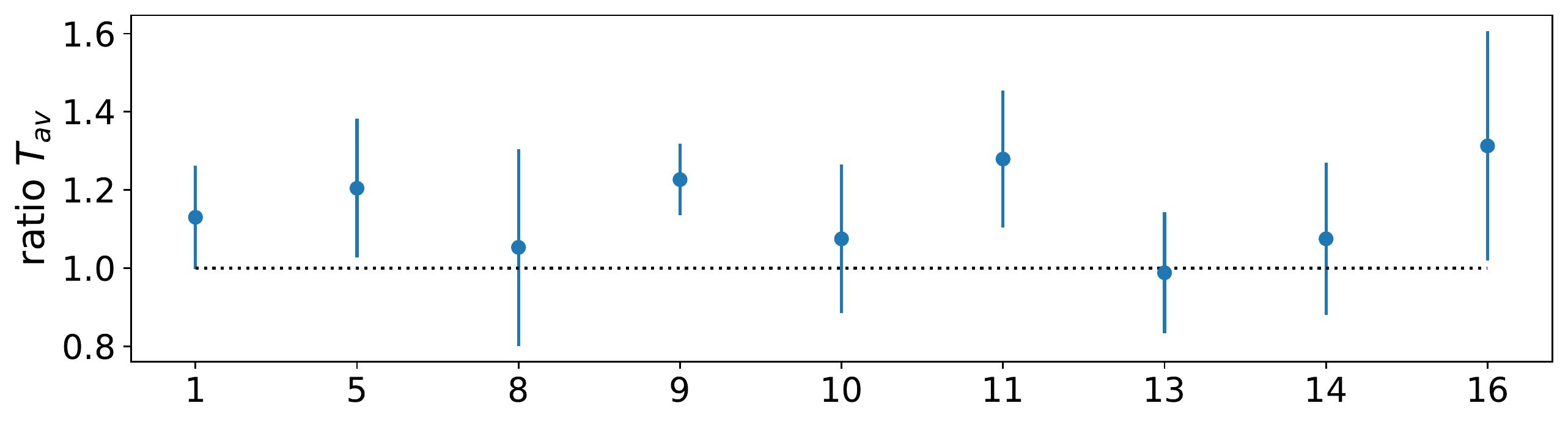}
    \caption{Parameter ratio of the flaring and quiescent state for all EPIC/pn light curves flagged variable with the changepoint analysis; see Sect.~\ref{subsubsect:analysis_epic_spectrum} for the definition of the two brightness states and Table~\ref{tab:QF_v} for the numbering of the observations.}
    \label{fig:flvsq}
\end{figure}

The changes in coronal flux and average thermal energy within each observation derived from the
time-resolved spectral fitting are visualised in Fig.~\ref{fig:flvsq}, where for each of the observations with a variable light curve, we display the ratio of the
highest and lowest brightness state. The errors were calculated with 
the error propagation for the values for quiescent and flaring part given in
Table~\ref{tab:QF_v}.

 Fig.~\ref{fig:flvsq} shows that $kT_{\rm av}$ 
significantly
changed 
  within the observations of July 2016 (observation 5), August 2018 (observation 9),
  August 2019 (observation 11), and January 2022 (observation 16). Out of these, observations 5, 11, and 16
  also show variations of the HR that are compatible with the 
variability 
identified in the count rate (see Fig. \ref{fig:hr_lc}). We conclude that all four observations
are likely to be affected by flare-like activity. 

Based on this variability analysis, we defined three
  different X-ray flux data sets for the subsequent analysis. X-ray flux set I contains only average fluxes
  from Table\,2 of \citet{Coffaro2020} and Table \ref{tab:epicpn_spectralparams}, 
  which means that no correction was applied for
  possible flaring activity. In X-ray flux set II, the average fluxes for all observations
  that are flagged as variable are replaced by the quiescent fluxes from Table \ref{tab:QF_v}. 
  In X-ray flux set III, only the four measurements with significant thermal energy variation described above are replaced by the respective quiescent fluxes.

\begin{table}[]
\caption{{\label{tab:QF_v}} X-ray flux and emission-weighted average thermal energy for quiescent and flaring time-intervals of the EPIC/pn observations of $\epsilon$\,Eri that show variability (see text in Sect.~\ref{subsubsect:analysis_epic_spectrum})}.
\tiny
\begin{tabular}{cccccc}
  \hline
  \hline
\noalign{\smallskip}  
No.  & Obs & \multicolumn{2}{c}{$F_{\rm X}$}                & \multicolumn{2}{c}{$kT_{\rm av}$} \\
obs.&    & \multicolumn{2}{c}{[$10^{-11}$\,erg\,cm$^{-2}$\,s$^{-1}$]} & \multicolumn{2}{c}{{[keV]}}     \\
&    & Quiescent               & Flaring              & Quiescent        & Flaring        \\ 
\hline
\noalign{\smallskip}
1 &02/2003      &$      1.23    \pm     0.01    $&$     1.47    \pm     0.02    $&$     0.36    \pm     0.02    $&$     0.40    \pm     0.05    $\\
5 & 07/2016     &$      1.76    \pm     0.03    $&$     2.03    \pm     0.02    $&$     0.39    \pm     0.03    $&$     0.46    \pm     0.06    $\\
8 & 01/2018     &$      1.80    \pm     0.04    $&$     2.05    \pm     0.02    $&$     0.42    \pm     0.09    $&$     0.44    \pm     0.05    $\\
9 & 08/2018     &$      1.86    \pm     0.02    $&$     2.37    \pm     0.02    $&$     0.38    \pm     0.02    $&$     0.47    \pm     0.03    $\\
10 & 01/2019    &$      1.20    \pm     0.01    $&$     1.33    \pm     0.03    $&$     0.32    \pm     0.02    $&$     0.34    \pm     0.08    $\\
11 & 08/2019    &$      1.56    \pm     0.02    $&$     2.08    \pm     0.02    $&$     0.41    \pm     0.05    $&$     0.53    \pm     0.04    $\\
13 & 08/2020    &$      1.44    \pm     0.02    $&$     1.73    \pm     0.04    $&$     0.36    \pm     0.03    $&$     0.36    \pm     0.06    $\\
14 & 01/2021    &$      1.71    \pm     0.02    $&$     1.83    \pm     0.03    $&$     0.39    \pm     0.03    $&$     0.42    \pm     0.07    $\\
16 & 01/2022 & $1.22 \pm 0.01 $ & $1.52 \pm 0.02$ & $0.33 \pm 0.02$ & $0.44 \pm 0.09$ \\
\hline
\end{tabular}\\
Notes: $F_{\rm X}$ and $kT_{\rm av}$ were derived from spectral fitting of EPIC/pn
      data for the light
      curves flagged as variable with the changepoint analysis;
      see Sect.~\ref{subsubsect:analysis_epic_spectrum} for the definition of the
      two brightness states. Errors are 95\% confidence levels computed with the
      XSPEC {\sc error} command.
\end{table}

\section{Results}

\subsection{Correlation of the data}

First, we investigated the correlation between the optical indices, \mws\, and \scairt ,
and the X-ray fluxes. These correlations have been examined on a statistical basis in
large stellar samples for example by \citet{Martinez-Arnaiz2011, Stelzer2013,  Martin2017, Mittag2017}.

In these studies, a single data point (mostly not contemporaneous for the different activity indicators that are compared) was  usually available for a given star.
Differently from this work, we searched for correlations between optical and X-ray activity diagnostics in 
a large data set for a single star. 

\begin{figure}
\begin{center}
\includegraphics[width=0.5\textwidth, clip]{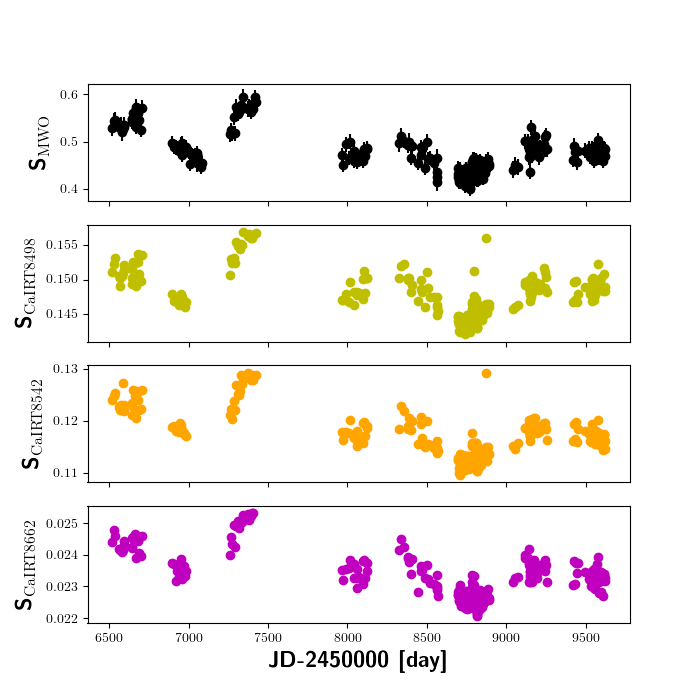}\\
\caption{\label{fig:ca_data} Time series of chromospheric indicators. TIGRE \mws\, (black dots; \textit{top}) and \scairt\,
  data (yellow, orange, and magenta dots corresponding to the lines at 8498, 8542, and 8662 \AA;\, \textit{lower three panels}).}
\end{center}
\end{figure}

For the optical indices, we only considered the TIGRE data because \scairt\ measurements were obtained only for these data. We compare
the \mws\, and \scairt\ data in
Fig. \ref{fig:ca_data}, where a correlation is apparent. 
We list all formal correlation coefficients in Table~\ref{tab:correl}. 
The Pearson correlation coefficient $r$ shows a very good correlation between these four line
indices with $r > 0.85$ and a probability $p$ that the correlation were zero of $p < 10^{-75}$  for all line combinations, except for
S$_{\rm MWO}$ to \scairt\ for the 
$8542\,{\rm \AA}$
line, which only has $r=0.81$. 

 \begin{table}
        \caption{\label{tab:correl} Pearson correlation coefficients for different chromospheric and coronal activity indicators. }
\footnotesize
\begin{tabular}[h!]{lcc}
\hline
\hline
\noalign{\smallskip}
    indicators & $r$ & $p$   \\
    \noalign{\smallskip}
    \hline
    \noalign{\smallskip}
\mws\,-- \scairt 8498 & 0.85  & 1.3$\cdot 10^{-91}$ \\
\mws\,-- \scairt 8542 & 0.81 &  1.3$\cdot 10^{-75}$\\
\mws\,-- \scairt 8662 & 0.88 &  8.3$\cdot 10^{-104}$\\
\scairt 8498 -- \scairt 8542 & 0.92 & 3.4$\cdot 10^{-129}$\\
\scairt 8498 -- \scairt 8662 & 0.90 & 3.6$\cdot 10^{-118}$\\
\scairt 8542 -- \scairt 8662 & 0.91 & 2.8$\cdot 10^{-126}$\\
mean \mws\,-- $F_{\rm X}$ set I ($T_{\rm diff}=20$ days) & 0.40 & 0.18\\
lowest \mws\,-- $F_{\rm X}$ set II ($T_{\rm diff}=20$ days) & 0.59 & 0.04\\

\noalign{\smallskip}
\hline

\end{tabular}
\normalsize
\end{table}

\begin{figure}
\begin{center}
\includegraphics[width=0.5\textwidth, clip]{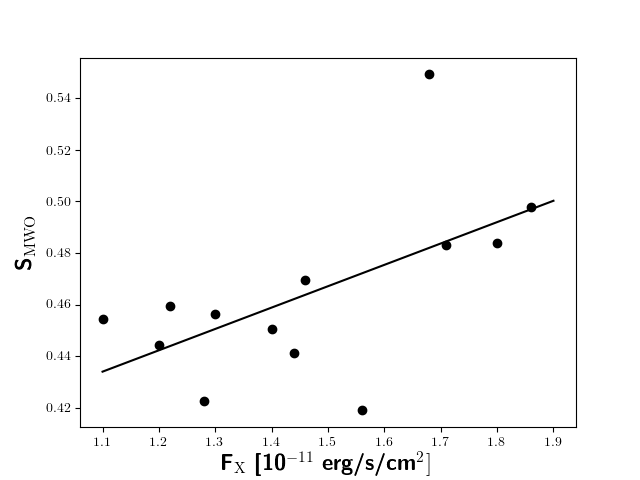}\\
\caption{\label{fig:corr} Correlation of most quiescent X-ray flux
        (set II of the X-ray fluxes, as defined in Sect. \ref{subsubsect:analysis_epic_spectrum}) and lowest
  \mws\ measurements within $T_{diff}=20$ days. Additionally, we
  show the best linear fit.}
\end{center}
\end{figure}

Next, we 
searched for a
correlation between the chromospheric \mws\ measurements and the coronal
X-ray flux measurements. For this comparison, the values should have been obtained 
truly simultaneously in the best case because for a growing time span between the observations, we  also expect a growing veiling of the correlation by short-term activity in both
X-ray and optical indicators \citep{Fuhrmeister2022}. 
We therefore constructed a \mws\, data set
considering only data within a certain time span $T_{\rm diff}= {\rm abs}(T_{\rm obs~XMM} - T_{\rm obs~Ca})$
of the \textit{XMM-Newton} observations. 
The shortest $T_{\rm diff}$ for which all X-ray observations have quasi-simultaneous optical observations is $T_{\rm diff}=90$ \,d. Since this time interval is too long for our purpose, we decided to opt for the shortest time interval that leads to a loss of only a small number of X-ray measurements,  
namely $T_{\rm diff}=20$\,d. For this choice, we had to exclude three \textit{XMM-Newton} observations,
because
no \mws\ observations are within $T_{\rm diff}$. We did not consider even lower values of $T_{\rm diff}$
because this would have left us with 
too few X-ray observations with quasi-simultaneous optical data. Since the rotation period of  $\epsilon$\,Eri of  $11.1-11.5$~d \citep{Baliunas1995, Froehlich2007}
is 
about half $T_{\rm diff}$,  the rotational modulation is expected to have been removed in the time-average values of \mws. X-ray time-series of active stars very rarely display rotational modulation. The short-term X-ray light curves often comprise irregular flare variability. However, since many flares last shorter than our chosen value for $T_{\rm diff}$
, a correlation between Ca\,II and X-ray emission is only expected if the amplitude of the activity cycle
is dominating the average activity level.
To mediate residual contributions from (flare-like)
short time variability,
we took only the quiescent, that is, the 
lowest values in both X-ray flux and \mws\ , data into account.  
Namely, we used the $F_{\rm X}$
values from X-ray flux set\,II as defined in
Sect.~\ref{subsubsect:analysis_epic_spectrum} (where all variable observations are
exchanged against quiescent flux) and 
the lowest  \mws\ value in each $T_{\rm diff}$ interval.
This should both roughly correspond to times at which the lowest number of active regions in chromosphere and corona are present on the visible hemisphere.
 We obtain a
 Pearson's correlation with $r=0.59$ and $p=0.04$, 
(listed in Table~\ref{tab:correl}. The correlation is  illustrated  
in Fig.~\ref{fig:corr}. 
When we instead use 
the mean X-ray flux (X-ray flux set\,I)
and the mean \mws\ value in the $T_{\rm diff}$ interval,
a much weaker and less significant
correlation is found,
with $r=0.40$ and $p=0.18$. 
This shows that there is a link between the  long-term variation of Ca\,II and X-ray emission.

\subsection{Cyclic activity behaviour}

 To access the cyclic activity, we employed the generalised Lomb-Scargle (GLS) periodogram \citep{Zechmeister2009, Scargle1982, Lomb1976} as implemented in PyAstronomy\footnote{\tt https://github.com/sczesla/PyAstronomy}
\citep{Czesla2019}. 

To allow a better visualisation and comparison of the data,
we also applied a  sinusoidal fit. We fitted the chromospheric activity indices (\mws, \scairt) as a function of time with sine waves; amplitude ($A$), period ($P$), phase ($ph$), and baseline value (y-axis offset) were free parameters, and we used the period corresponding to the highest peak in the
GLS as starting value for the fit.
We caution that cycles may show asymmetries, as is observed on the Sun, for example, where the cycle is better
described by a skewed Gaussian \citep{Du2011} 
because the rise is faster than the decay. However, 
 \citet{Willamo2020} found that the solar cycle is particularly asymmetric compared to stellar cycles.
 A single chromospheric cycle of
\epsEri shows 
some evidence for skewness, but a 
description with a sinusoidal curve nevertheless fits the data fairly well. 

\subsubsection{Detection of the known 3-year cycle in the X-ray data}

In the previous study of the X-ray cycle by \citet{Coffaro2020},
only the agreement with the short S-index cycle has been reported owing to the combination of sparse data coverage and short time baseline, and no  actual search for a periodicity in the X-ray data has been performed.
 Using all X-ray flux measurements and their errors from the timing analysis,
we are now in the position to perform a GLS analysis. This resulted in a formal detection of the 3-year cycle for the first time. 

\begin{figure}
\begin{center}
\includegraphics[width=0.5\textwidth, clip]{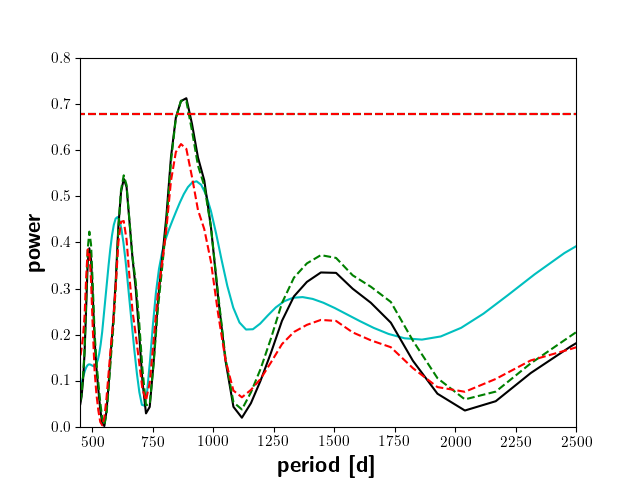}\\
\caption{\label{fig:glsxray} GLS power of the X-ray flux time series.
        The solid black line corresponds to the GLS power computed from the X-ray fluxes
        of  set\,III defined in Sect. \ref{subsubsect:analysis_epic_spectrum}, the dashed green line corresponds to 
  set\,I, and the
        dashed red line corresponds to 
        set\,II. The solid cyan line shows the power of the GLS for the
  TIGRE \mws\ measurements, which are observed in about the same time interval as the X-ray data. The dashed red line shows the 1\% FAP level for the
  X-ray GLSs.}

\end{center}
\end{figure}

The GLS periodogram is shown in Fig.~\ref{fig:glsxray}.
We wished to avoid contribution from obvious flaring activity, but on the other hand, we did not wish to cut longer-lasting higher activity states. We therefore used X-ray flux  set\,III  as defined in Sect.~\ref{subsubsect:analysis_epic_spectrum}
and determined a period of 881.33 $\pm$ 11.4\,days ($\sim$ 2.4 $\pm$ 0.03\,yr).  To obtain the error, we simulated 10000 data sets of the X-ray flux values. Each data point was randomly drawn from a normal distribution within the standard deviation around the measured $F_{\rm X}$. The standard deviation of the obtained periods from each simulated data set was then considered as the error of
the X-ray period. 

Although the cycle length from the X-ray flux is somewhat shorter than previously published values from chromospheric indicators,
it roughly agrees with the cycle length
determined from  chromospheric data taken during 
the  approximate time-span covered
by the {\it XMM-Newton}  observations.  This value can be found together with
a more detailed discussion of the timing behaviour of the cycle length in  Sect. \ref{knowncyc}.

\subsubsection{Evidence for a $34$-year activity cycle from \mws\ data}

 The results from the GLS analysis of the  S-index
are shown in Fig. \ref{fig:gls}. The highest peak occurs at a rather
long period of $12355 \pm 230$\,days ($\sim$33.8\,yr) for the whole data set (the error was obtained in the same way as for the X-ray data).

This 
$34$-year period has also been found in the Mount Wilson program data, but was discarded by
\citet{Metcalfe2013} 
because it was similar in length to the data set. 
Consistent with the result by  \citet{Metcalfe2013},
we also find this
period in the Mount Wilson program data alone, but not in the TIGRE data alone 
 because its
time baseline is not long enough. 

\begin{figure*}
\begin{center}
  \includegraphics[width=1.03\textwidth, clip]{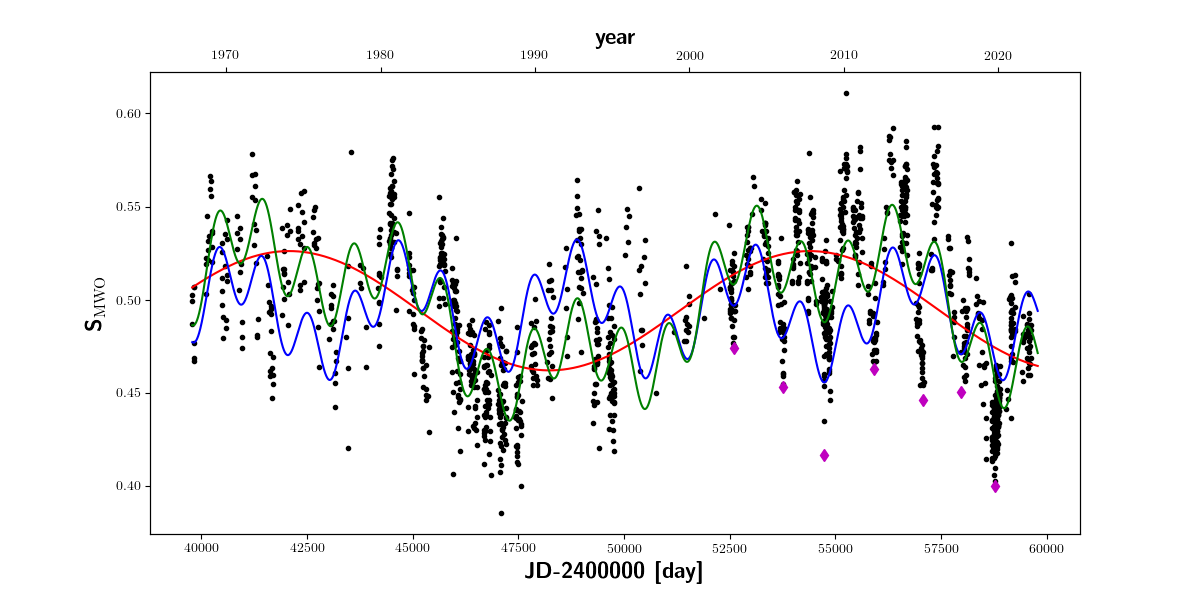}\\
  \includegraphics[width=1.03\textwidth, clip]{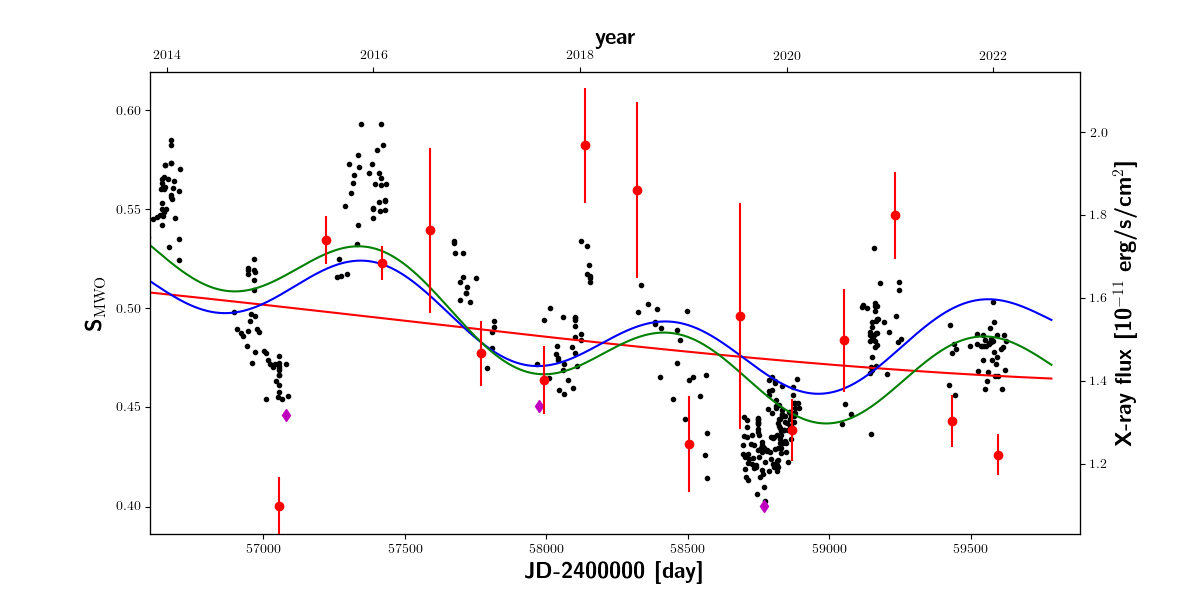}\\
\caption{\label{fig:timeseries} Time series of the \mws\ data (black dots).  The best
  sine fit of the \mws\ data with one period is indicated by the  solid red line, and the solid blue and green lines indicate the best sine fit with two and three periods, respectively.
  {\textit Top:} All data. {\textit Bottom:} Zoom into the top panel for the most recent decade. The $F_{\rm X}$ measurements from set III (red filled circles) with their errors are also plotted. In both panels,
  the magenta diamonds mark the data points  of the minimum \mws\ value in each cycle. 
}
\end{center}
\end{figure*}

 We show the whole time series in Fig.~\ref{fig:timeseries}. It 
still covers less than two
periods. Nevertheless, while the data available to
\citet{Metcalfe2013} roughly extended from one maximum to the next,
the more recent
data show that the decay phase follows the second maximum. This adds strong evidence for a long-term  periodicity. However, it is not yet clear  whether this period is caused by a magnetic cycle. \citet{Jeffers2022} found a switch in the signed average magnetic field
of \epsEri at about 2007 that approximately coincides with 
the maximum of our long period. This would be in line with findings for the Sun \citep{Sanderson2003} and 61 Cygni A \citep{BoroSaikia2016, BoroSaikia2018a}, where the
reversal of the magnetic field was also found to occur around the activity maximum. Further measurements are needed to confirm whether this coincidence of magnetic field reversal
and cycle maximum in \epsEri is caused by a magnetic cycle or occurred by chance.
For the Sun, cycles longer than the Schwabe cycle (e.\,g. the 90-year Gleissberg cycle and the 210-year de Vries cycle) have also been discussed to
be caused by noise \citep{Cameron2019}. Nevertheless, the long cycle explains the observed decay of
the \mws\ values between $\sim$2016-2022.   To illustrate this, we performed a series of sine-curve fits on the whole \mws\ data set. All fitting parameters can be found in Table~\ref{Tab:sinfit}. 

 First, we fitted the data with a single sine-curve
for the longest period found using its length from the GLS analysis as input parameter for the fit. This resulted in a period of $12311$\,days. 
Fitting instead with the two known and well-established shorter cycle periods as input parameters,
this leaves us with periods of $3980$ and $1061$\,d. For a fit with three free periods, we obtain best-fitting periods of $12578$, $3960$, and $1062$\,d. These numbers all agree with the cycle lengths 
obtained from the GLS analysis discussed in Sect. \ref{knowncyc}. A comparison of the standard deviations of the 
fits with two and with three periods
($0.035$ and $0.027$, respectively) shows that the 
latter period
reduces the scatter in the residuals and
is therefore a better description.
We refrained from computing a reduced $\chi^{2}$ because no errors are assigned to the Mount Wilson data. We show the fits with one, two, and three cycles together with the \mws\ data in Fig.~\ref{fig:timeseries}.
In the bottom panel of Fig. \ref{fig:timeseries}, we show a zoom-in for the last nine years, also including the coronal X-ray data. 

\subsubsection{Known 3-year  and 12-year  activity cycles}\label{knowncyc}

The periodogram also shows peaks at 3970 $\pm$ 54\,days ($\sim$10.9\,yr) and at
1060$\pm$30\,days ($\sim$2.9\,yr).
These two periods have been proposed as activity cycles
before, as discussed in Sect.~\ref{sec:intro}, when we ascribe the longer of our periods to the 12.7 yr cycle found by \citet{Metcalfe2013}. 
Both periods are also found  
when using the Mount Wilson program data or TIGRE data alone. 

A high peak lies at about half of the 3 yr period; this can be identified in every data set. This peak 
corresponds to the alias of the 3 yr period and the one-year observation pattern. 
Two peaks also flank the
peak at 11 yr ($\sim$3970 days), which can be identified as aliases of the 34 yr period. 

We also
computed a GLS in which the best-fit sine corresponded to a subtracted 34 year period, which led to a damping of these aliases,
but not of the 11 yr period. Moreover, the peak of the 3 yr (=1060 days) period is then strongly enhanced. The formal false-alarm probability (FAP) for the 34-year period 
and the FAPs of the 11 and 3 yr period with the 34 yr period
removed are all lower than $10^{-10}$. We caution, however, that the FAPs
are probably underestimated because of the high number of observations.

\begin{figure}
\begin{center}
\includegraphics[width=0.5\textwidth, clip]{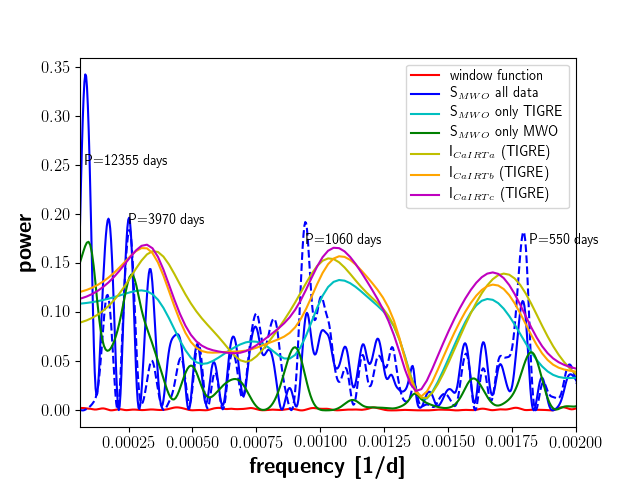}\\
\caption{\label{fig:gls} GLS power of the whole \mws\ time series (solid blue line)
  and the whole time series without the longest period (dashed blue line).
  We also show the GLS power of the window function (red line), of the
  \mws\ data of the TIGRE telescope and the Mount Wilson program data alone (cyan and
  green line, respectively), and of the \scairt\ of the three \ion{Ca}{ii} IRT lines (yellow, orange, and
  magenta with increasing central wavelength). 
}
\end{center}
\end{figure}

Furthermore, we used  fits of sine-curves for a comparison between \mws\ and \scairt\ data.
We applied a fit of a single sine function representing the 3-year cycle to the TIGRE data (amplitude, period, and phase as free parameters; the initial values were set to the GLS result) and obtained a cycle length of 928 days for \mws\ data and 959, 923, and 932 days for the
three \scairt\ data sets listed  here in order of increasing line wavelength.
The periods from the sine fits agree within 2$\sigma$ with the period value obtained from the GLS  analysis of the \mws\ TIGRE data. 
The recovery of the $3$-year period from the GLS  periodogram (see Fig.~\ref{fig:gls}) in the sine-fits lends credibility to the detection of the $3$-year cycle 
in the \scairt\ data.

Next to the cycle length, we also compared the amplitudes of the different sine fits, $A$,
but we caution that the sine fits underestimate the amplitude systematically, as can be seen in Fig.~\ref{fig:timeseries}.
We found that the sine-fit amplitude is highest for the \mws\ TIGRE data set with $A=0.034$ ($A$/offset  = 7\%), while
the three \scairt\ data sets lead to amplitudes of $A=0.0030$ (2\%), $A=0.0039$ (3\%), and $A=0.0007$ (3\%). This means that the reddest \ion{Ca}{ii} IRT line is the hardest to measure because the absolute magnitude is lowest. All three \ion{Ca}{ii} IRT lines are comparably sensitive to the cycle because the relative amplitude ($A$/offset) is about the same. However,
we find a higher sensitivity of the \ion{Ca}{ii} H\&K lines to the
cyclic activity variations. We list all fitting parameters of our sine fits in Table~\ref{Tab:sinfit}.

Finally, we examined the cycle length variation in the 3-year cycle, which is quite
evident by the generally shorter cycle length obtained from TIGRE \mws\ data compared to the whole \mws\ data set (1061.91 days vs. 927.9 days; see Table \ref{Tab:sinfit}). We performed a GLS search in 6-year-long time intervals and report the measured cycle length and the time intervals in Table \ref{Tab:cyclevar}. We find a
minimum and maximum cycle length of 910 and 1350 days, respectively, which roughly corresponds to the relative cycle-length variations of the Sun, that is, 8--15 years \citep{Richards2009}. There is no evident pattern that would reveal a systematic variation of the cycle length.  

\begin{table}
        \caption{\label{Tab:sinfit}  Best-fit parameters obtained by fitting the \mws\ and \scairt\ data with sine curves ($A\cdot \sin(2\cdot\pi/P\cdot time +ph)+$offset)}
\tiny
\begin{tabular}[h!]{lcccc}
\hline
\hline
\noalign{\smallskip}
 Fit & $A$ & $P$ & $ph$ & offset \\
  &  & [day] &  &  \\
\noalign{\smallskip}
\hline
\noalign{\smallskip}

\mws\, only 33yr period & 0.03 & 12311.1 & -63.5 & 0.49 \\
\mws\, 2 sine curves & 0.021& 3984.118 & -6.273 & 0.49\\
                     & 0.017 & 1061.453& 1.1665 & \\
\mws\, 3 sine curves & -0.0317 & 12578.057 & 60.247 & 0.49 \\
                     & 0.0187 & 3960.685 & 12.408 &  \\
                     & -0.0191  & 1061.911 & 29.052 &  \\
\mws\, TIGRE, 3yr cycle & -0.034  & 927.9 & -1197.2 & 0.47 \\
\scairt$_{a}^{a}$\, 3yr cycle & -0.0030 & 958.7 & -662.7 & 0.15 \\
 \scairt$_{b}$\, 3yr cycle& -0.0039 & 922.7 & -1291.3 & 0.12 \\
\scairt$_{c}$\, 3yr cycle & -0.0007 & 932.4 & -1117.9 & 0.02 \\
\noalign{\smallskip}
\hline

\end{tabular}\\
\normalsize
Notes: $^{a}$ Indices a, b, and c refer to the individual lines of the triplet in the order of increasing wavelength.
\end{table}

\begin{table}
        \caption{\label{Tab:cyclevar}  Length of the short (3-year) cycle in subsequent 6-year long data intervals determined from the highest GLS peak or minimum to minimum measurements.}
\tiny
\begin{tabular}[h!]{cccc|cc}
\hline
\hline
\noalign{\smallskip}
 JD first & JD last & no.  & period & min-min$^{a}$&ampl.$^{b}$ \\
 --2400000 & --2400000& spec & &  & \\
 $[$day$]$ & [day] & & [day] & [day] & \\
\noalign{\smallskip}
\hline
\noalign{\smallskip}

 39786.8 & 41976.8 & 87  &  931.7 \\
 41977.8 & 44167.8 & 64  &  no peak\\
44168.8 & 46358.8 & 202 &  1128.7\\
 46359.8 & 48549.8 & 220 &  no peak\\
 48550.8 & 50740.8 & 133 &  1354.9\\
 50741.8 & 52931.8 &  66 &  (903.3)$^{a}$\\
 52932.8 & 55122.8 & 265 &  1091.5 & 1148, 966 & 0.041, 0.043\\
 55123.8 & 57313.8 & 231 &  1124.5 & 1185, 1171& 0.082, 0.058\\
 57314.8 & 59504.8 & 271 &  909.6 & 895, 796& 0.044, 0.079\\

\noalign{\smallskip}
\hline
\multicolumn{3}{l}{mean cycle length:} & 1063$\pm$151\\

\noalign{\smallskip}
\hline

\end{tabular}\\
\normalsize
Notes: 
$^{a}$ Does not fulfill the significance level of FAP $<$ 0.1\%.
\end{table}



\subsubsection{Length-to-amplitude law}

For the Sun, a length-to-amplitude law for adjacent activity cycles is known
\citep{Hathaway2015} that states an anti-correlation between
the cycle length and the amplitude of the subsequent 
cycle. \citet{Hathaway1994} and \citet{Solanki2002}, who studied this relation, used Sun spot number and not S-index measurements, but because these two
are highly correlated, the length-to-amplitude law should hold for the S-index as well, even though the minima of these different
cycle indicators are shifted slightly with respect to each other for the Sun. Moreover, the Waldmeier effect which was first known from Sun spot number as well has been shown to be present also in the solar S-index data \citep{Garg2019}. 

Cycle length and amplitude can be measured for \epsEri for several adjacent $3$-year cycles. Fig.~\ref{fig:timeseries} all cycle minima of the
$3$-year cycle after 2002 can be identified by eye. 
Before 2002, the cycle minima can only partly be identified.
For some cycles, 
this is a result of sparser sampling, but in some cases, 
 the cycles were less pronounced. 

Analogously to the case of the solar cycle,
we measured the cycle length as the time between consecutive minima (as defined by the lowest \mws\
measurements; these are highlighted
in Fig.~\ref{fig:timeseries}).  We then computed the median \mws\ value of each cycle and subtracted it from
the mean of the highest three \mws\ values in the respective cycle. In this way,
we derived the amplitude of the cycle without relying on a single measurement (which may be affected by flaring)
and also corrected for the $34$-year cycle. We used the median \mws\ value as a reference here and not the minimum because the next minimum may differ significantly, so that start
and end of a cycle would not have the same activity level as measured by \mws.
 We took these measurements on the last three 6-year time intervals defined in Sect.~\ref{knowncyc}, which contain two cycles each.
We list the values for the cycle length and their amplitude
for the six $3$-year cycles after 2002 in Table~\ref{Tab:cyclevar}.  
Since we have length and amplitude measurements for six cycles, we have five
data pairs (length of the n-th cycle and amplitude of  cycle n+1) according to the length-to-amplitude law for the Sun.
 
We find an
anti-correlation for the cycle length and the amplitude of the subsequent
cycle
in \epsEri, with  a Pearson correlation coefficient of
$r=-0.89$ and a probability $p=0.04$. Although
the anti-correlation test is based on only five data points, 
it  suggests  comparable
laws for \epsEri and the Sun.

\subsection{Rotation period from optical measurements}

The rotation period of \epsEri cannot be found in the whole
dataset of the \mws\ data with a GLS analysis: No outstanding peak can be identified in the GLS in the range between 10 and
13 days. However, when only MWO data are used, there
is an outstanding peak at about 11.1\,d, but with an FAP > 0.1. In the TIGRE data, the rotation period can also be identified in the \scairt\ data with an 
FAP lower than 3\% for all three line indices and with an FAP$<$1\% for the bluest
line without detrending the data. The highest peak for all three
\ion{Ca}{ii} IRT lines is at a rotation period of 11.8 days, 
which agrees well with
the periods found by \citet{Froehlich2007} (photometry) and \citet{Baliunas1995} (\mws measurements). 
The TIGRE \mws\ data instead show that the FAP of the peak
at the rotation period using the \mws\ data is again much worse, but it is still
the highest peak. This hiding of the rotation period can first be caused by the longer trends in the data sets (with \mws\ having a higher
cycle amplitude than \scairt), or second, by differential rotation (e.g. the period difference to the MWO data set suggests). A third possibility is the evolution of 
individual bright plage regions on timescales shorter than the rotation period of
\epsEri. Again the \mws\ data would be affected more strongly because these lines are more sensitive to changes in activity than \scairt. Strong changes in \mws\ values
of \epsEri on timescales shorter than the rotation period have been found \citep{Petit2021}. Evolution of activity features on short timescales is also known in a different context, for instance, the evolution of prominences as found for V530~Per \citep{Cang2020}.
\begin{figure}
\begin{center}
\includegraphics[width=0.5\textwidth,clip]{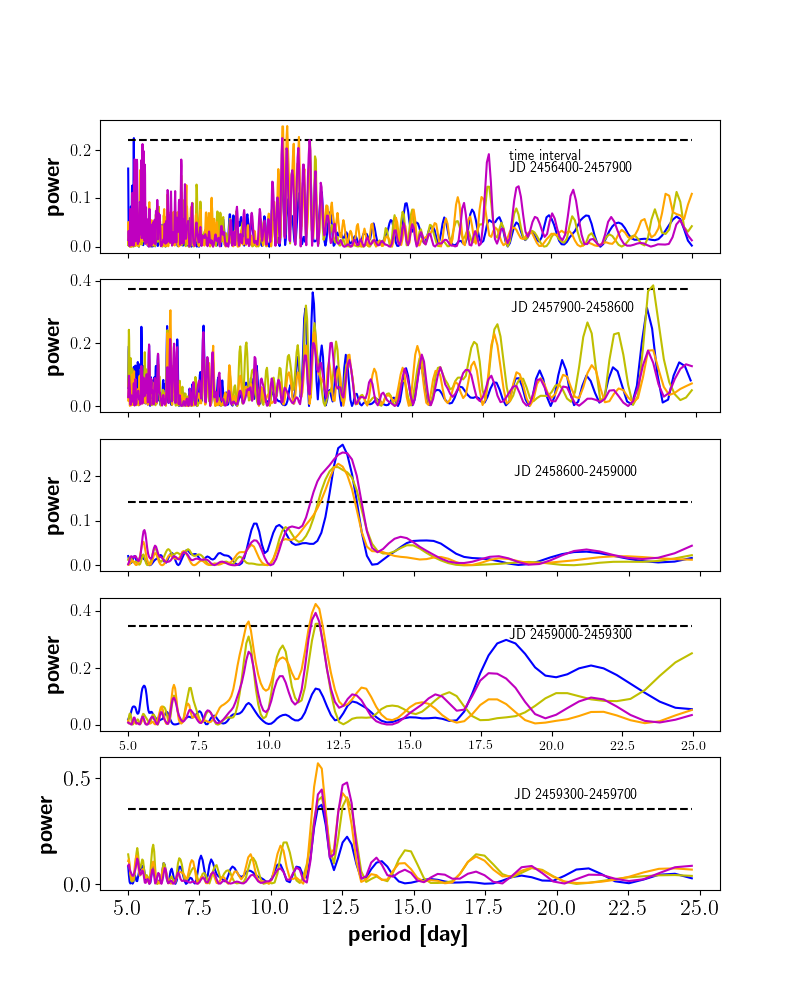}\\
\caption{\label{fig:rotper} GLS power of the \scairt\ time series for the
  \ion{Ca}{ii} IRT lines (yellow, orange, and magenta, with increasing central wavelength)
  and of the TIGRE \mws\ measurements (blue) for different observation seasons,
  see Table~\ref{tab:rot}. The dashed horizontal line marks the
  FAP $<$0.05. 
 }
\end{center}
\end{figure}
%

To reduce the influence of long-term variations,
we computed the rotation period in five shorter time intervals that roughly corresponded to one to two TIGRE observing seasons depending on
the number of observations. 
We list the 
start and end
of
each time span, the corresponding number of observations, and
the period obtained from the highest GLS peak for each time interval
and each of the four chromospheric activity  indicators  in Table~\ref{tab:rot}.
Using these shorter time spans, we subtracted a second-order polynomial 
from  the data of each time
interval and each index before computing the GLS periodogram, which we show in Fig.~\ref{fig:rotper}. The rotation
period  determined in this way shifts slightly for different observation seasons, which
may be caused by differential rotation.
For example we find a systematically higher rotation period of 12.3-12.5\,d in all lines in the third time interval compared to all other time intervals.
 \citet{Petit2021} found that differential rotation on \epsEri ranges from about 10.8 days for equatorial spot latitudes to 13.3 days for polar latitudes. Our measurements
all fall into this range, with the exception of one measurement and the few cases for which we find periods at (roughly) half, double, and $1.5$ times the 
rotation period for individual activity indicators and time spans (see Table
\ref{tab:rot}).  

A significant rotation period with an FAP $<$ 5\% is found for each index in the third to fifth time intervals, except for one time interval in the
\mws\ data. These time intervals roughly correspond to the late decay phase and minimum of the $34$-year activity cycle. In the first two time intervals, which correspond to the early decay phase of
the $34$-year  activity cycle,  we do not recover the rotation period with a significant FAP, though there is a comparable
number of data values available. This somewhat contra-intuitive finding may be explained by the generally high activity  of \epsEri.  
 It may be that  during the peak of the 34-year  maximum flaring or other types of activity, variations veil the variability pattern of plages rotating with the star.
It is also possible
that the filling factor of plages is too high to lead to a rotation pattern at activity maximum. This latter explanation is in line with the finding by \citet{Coffaro2020} regarding the corona of \epsEri. The corona was shown to be covered by up to $90$\,\% 
with
magnetically active structures, which explains the relatively low amplitude of the X-ray cycle.

An alternative explanation may be that the first two intervals cover long times (the first interval more than one, the second interval nearly one 3-year cycle), while the latter three intervals are better sampled in time and cover each about one-third of the 3-year cycle (compare Table \ref{tab:rot} and
Fig. \ref{fig:timeseries}). Changes in rotation period caused by differential rotation and changing spot latitudes may therefore dilute the detection
of a rotation period in the first two intervals. Interestingly, the third time interval corresponding to the 3-year cycle minimum shows the longest rotation period, while the intervals covering the rise phase and maximum of the cycle show a shorter period. This may indicate that a butterfly diagram can be found in \epsEri as well, with longer rotation periods (i.e. spots or plages at higher latitudes) at cycle minimum and a migration of active regions to more equatorial latitudes during the later cycle phases. Further observations with dense sampling to detect rotation periods for distinct phases of the cycle are clearly needed here.

\begin{table}
        \caption{\label{tab:rot} Rotation periods as determined from the
        given time intervals.}
\footnotesize
\begin{tabular}[h!]{cccccc}
\hline
\hline
\noalign{\smallskip}
   no. & min-max & period & period & period & period \\
   spec & JD &  \mws & \scairt$_{a}$  & \scairt$_{b}$ & \scairt$_{c}$ \\
  & -2450000   & [day] & [day] & [day] & [day]\\
 \noalign{\smallskip}
    \hline
    \noalign{\smallskip}
    53 & 6400--7900 &5.2$^{a}$  & (11.5) & 10.6 & (10.5)$^{b}$\\
    39 & 7900--8600 & (11.5) & 23.5 & (6.5) & (7.6)\\
    126&8600--9000 & 12.5  & 12.2 & 12.3 & 12.5\\
    42&9000--9300 & (18.4) & 11.8 & 11.6 & 11.6\\
    41&9300--9700 & 11.8 & 11.8 & 11.6 & 12.6\\
\noalign{\smallskip}
\hline

\end{tabular}\\
Note: $^{a}$ All values have errors of 0.1 or lower computed with a Monte Carlo simulation; $^{b}$ Values in brackets have an FAP > 5\%.
\normalsize
\end{table}

\section{Discussion and conclusions}

We revisited the chromospheric and coronal activity cycle of \epsEri\ using 
time
series of 
 Ca\,{\sc ii} emission in the H\&K lines and the IRT and of X-ray emission. The H\&K data originate
from different telescopes (including data from the
Mount Wilson program, the Lowell observatory, and the TIGRE telescope) covering more than $50$ years, the Ca\,IRT data cover 9 years, and the
X-ray flux measurements by {\it XMM-Newton} cover more than 7 years.
In both 
chromospheric and coronal emission,
we can establish the already known 
short ($\sim 3$\,yr)
activity cycle. With the analysis of the X-ray time series, we present
the first quantitative detection of this
cycle in X-rays, although \cite{Coffaro2020} reported 
X-ray fluxes 
from a shorter {\it XMM-Newton} data set previously, in agreement with the Ca\,H\&K  cycle.
Moreover, we 
detected
a medium-length activity cycle of about $3970$\,d ($\sim$10.9\,yr) in the chromospheric data, which is
considerably shorter than the $12.7$\,yr period found by \citet{Metcalfe2013}. This discrepancy
can be explained by the  
existence
of an even longer cycle with a period of $12355$\,d  ($\sim$33.8\,yr)
that reveals its presence only in the whole data set.

Additional longer cycles, next to the well-known Schwabe (11-year) cycle, have been proposed for the Sun as well. The most prominent  longer cycle is the Gleissberg ($\sim$80-100 years) cycle, which, interestingly, is roughly ten times longer than the Schwabe cycle, similar to the cycle ratio for   \epsEri ($3$ versus $34$\,yr). Several other longer and shorter periodicities have been found in proxies
of solar activity, such as radioisotope concentrations (see \citet{Usoskin2013}, \citet{Hathaway2015}, and references therein). The de Vries or Suess cycle, for
example, lasts $\sim$ 210 years \citep{Suess1980}, while an even longer $\sim$2400-year cycle was discussed by \citet{Damon1991}. 

While 
the long $34$\,yr period of \epsEri
is highly significant in the GLS, only 1.5
cycles are covered so far, and future observations are needed to verify whether this is only a quasi-periodic
episode or a true long-duration cycle. However, if it is true, this long cycle naturally
explains the drop in the \mws\ data in 2018, which was interpreted as a change in the cycle behaviour
by \citet{Coffaro2020}. 
 
Further properties of the solar cycle that we were able to examine on \epsEri based on the extraordinary long time series of chromospheric measurements are variations in cycle length.
The 11-year Schwabe cycle is well known to show variations of its length that may vary roughly between 8 and 15 years for individual cycles \citep{Richards2009}. For \epsEri, we investigated the short cycle for length variations and found it to be variable at a standard deviation of 151 days for the measurements of the period length, with the period ranging 
from $910$ to $1355$\,d, which is a fractional variation about as high as that of the solar $11$\,yr cycle.  Moreover, we tentatively also report a length-to-amplitude law
in our \mws\ data as is known for the Sun from sunspot numbers. Further adjacent well-defined cycles will clarify whether the length-to-amplitude law indeed holds for \epsEri as well.

Our long-term monitoring of $\epsilon$\,Eri also allowed us to estimate the long-term X-ray minimum state of the star by subtracting short-term variability as well as cycle variations. As representative for this  long-term averaged quiescent state we considered the five {\it XMM-Newton} observations with the lowest flux in the  quiescent segment of their EPIC/pn light curves (observations 1, 2, 10, 12, and 16), from which we obtain $1.21 \pm 0.06 \cdot 10^{-11}\,{\rm erg/cm^2/s}$. Comparing this to the X-ray luminosity function (XLF) of K dwarfs in \cite{Preibisch05.0}, we found that the X-ray luminosities of only $\sim 10$\,\% of the field K dwarfs are higher than $\epsilon$\,Eri in its lowest activity state. The activity level of $\epsilon$\,Eri is high for its spectral type and is most likely to be attributed to its young age. We can therefore conclude in reversing the argument that only $\sim 10$\,\% of the field K dwarfs are younger than $\sim 500$\,Myr, the age of $\epsilon$\,Eri. 
When we allow that the X-ray flux of most of the stars in the XLF presented by \cite{Preibisch05.0} may include contributions from  flares and cycles, 
many older stars may scatter into the upper $10$\,\% of the XLF, such that the fraction of young stars is likely lower than $10$\,\%. Improved XLFs using a {\it Gaia}-based census of the solar neighbourhood combined with updated X-ray data from the  eROSITA instrument \citep{Predehl21.0},  for instance, should be employed to verify this conclusion.

In summary, our multi-wavelength study provided further insight into the long-term activity of \epsEri: (i)
We determined the coronal X-ray 
cycle to be $881.33$\,d, in agreement
with the measurement of the $3$-year cycle from the TIGRE data 
in
the same time span as the X-ray observations. 
(ii) We presented strong evidence for a long 34-year activity cycle in \mws\ data. 
(iii) We demonstrated the detectability of the short and the medium cycle in \ion{Ca}{ii} IRT data for the first time. The cycles
give about the same values as the simultaneous \mws\ measurements, even though the amplitude
of the two cycles in the \ion{Ca}{ii} IRT data is lower. This is relevant in the context of
the {\it Gaia} spectra, which cover the \ion{Ca}{ii} IRT lines, but not the \ion{Ca}{ii} H\& K lines. Activity cycle detections should therefore be possible with 
 Gaia RVS spectra in principle.
(iv) We detected variations in cycle length of the 3-year and 11-year cycle that are
comparable to that of the Sun.
(v) We find a moderate correlation between X-ray flux and S-index measurements, but they are not simultaneous. 
This demonstrates that the
long-term variation of the activity level of \epsEri governs the quiescent emission of chromosphere and corona. 
(vi)  We established the long-term averaged minimum X-ray state of $\epsilon$\,Eri, which places the star among the $< 10$\,\% most active K dwarfs in the solar neighbourhood. (vii)
The previously known rotation period of $11.8$\,d  was found
in the \ion{Ca}{ii} IRT data, but not in the \mws\ data, which indicates that these lines should (additionally)
be used as indicators for rotational
period and activity cycles to confirm findings by \mws\ data or when no \mws\ data are available.

\begin{acknowledgements}
 M.C. acknowledges funding by Bundesministerium für Wirtschaft und Energie through the Deutsches Zentrum für Luft- und Raumfahrt e.V. (DLR) under  grant FKZ 50 OR 2008.
 We thank our referee Pascal Petit for helpful suggestions. The HK\_Project\_v1995\_NSO data derive from the Mount Wilson Observatory HK Project, which was supported by both public and private funds through the Carnegie Observatories,
 the Mount Wilson Institute, and the Harvard-Smithsonian Center for Astrophysics starting in 1966 and continuing for over 36 years. These data are the result of the dedicated work of O. Wilson, A. Vaughan, G. Preston, D. Duncan, S. Baliunas, and many others. TIGRE is a collaboration
of the Hamburger Sternwarte, the Universities of Hamburg, Guanajuato and
Liège. J.S.F. 
This work used data taken with the {\it XMM-Newton} X-ray space observatory
operated by the European Space Agency (ESA).
\end{acknowledgements}

\bibliography{papers}
\bibliographystyle{aa}

\begin{appendix}

\section{X-ray light curves}
\label{sec:ap2}

\begin{figure*}
\begin{center}
\includegraphics[width=0.5\textwidth,clip]{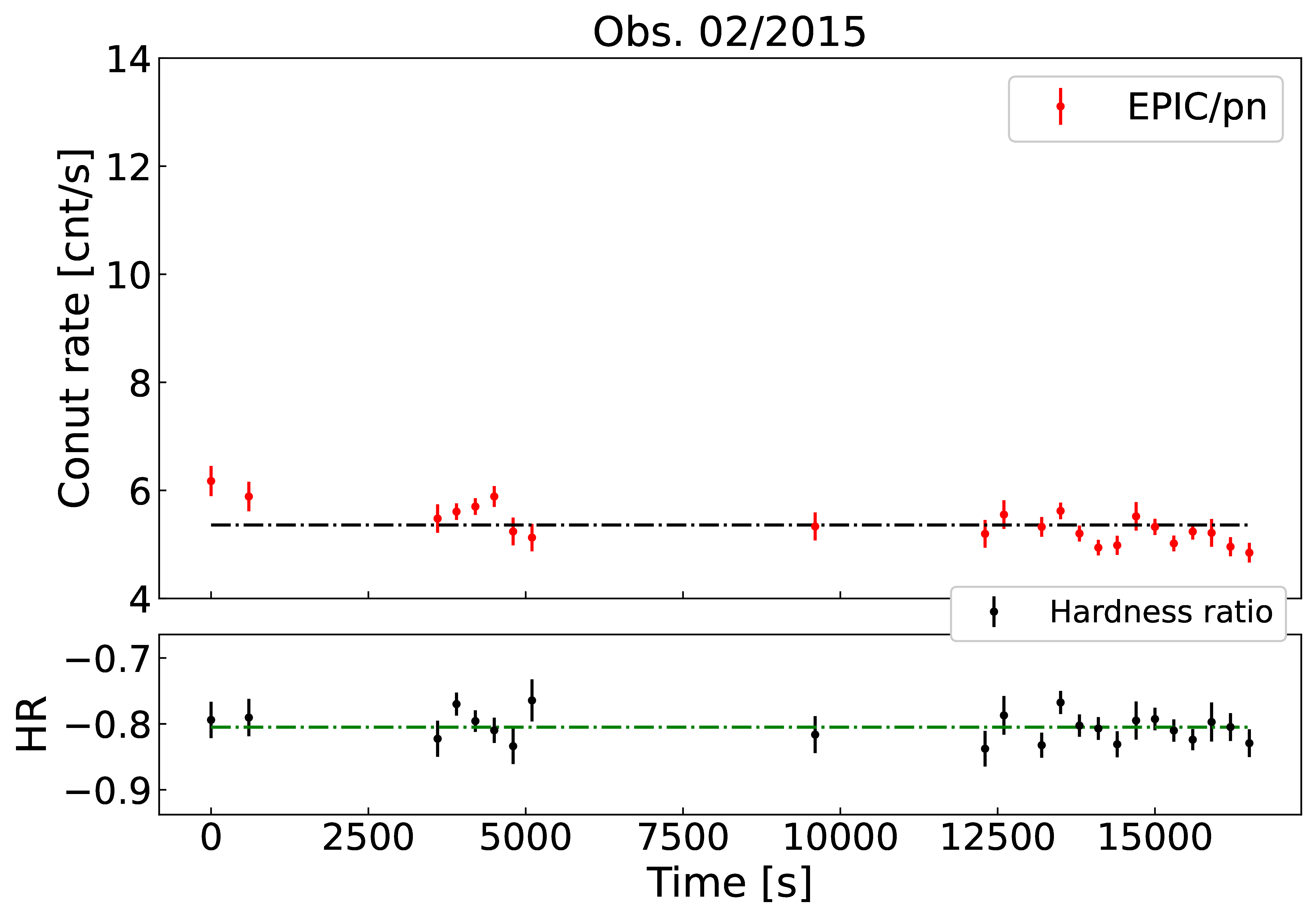}
\includegraphics[width=0.5\textwidth,clip]{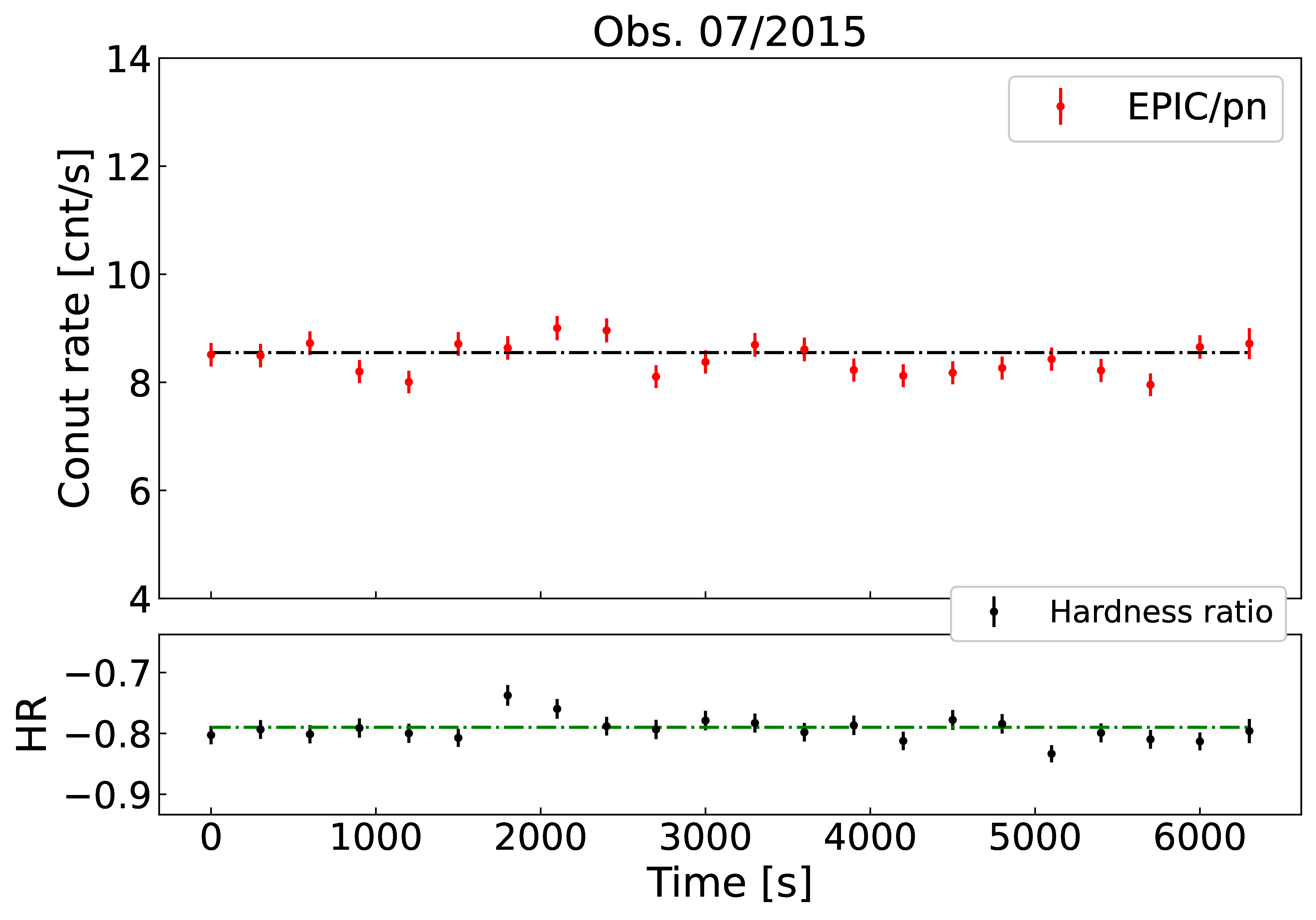}\\
\includegraphics[width=0.5\textwidth,clip]{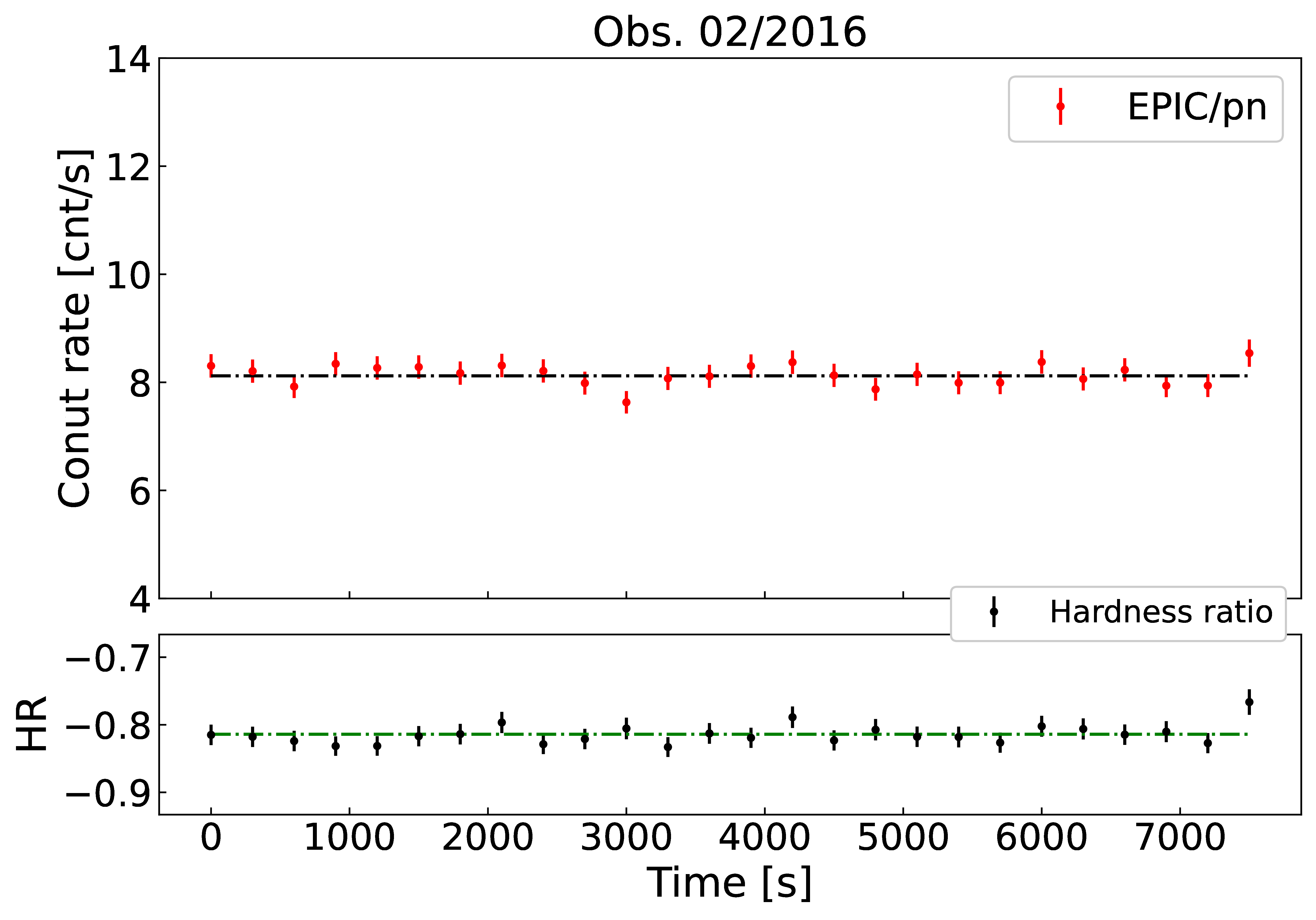}
\includegraphics[width=0.5\textwidth,clip]{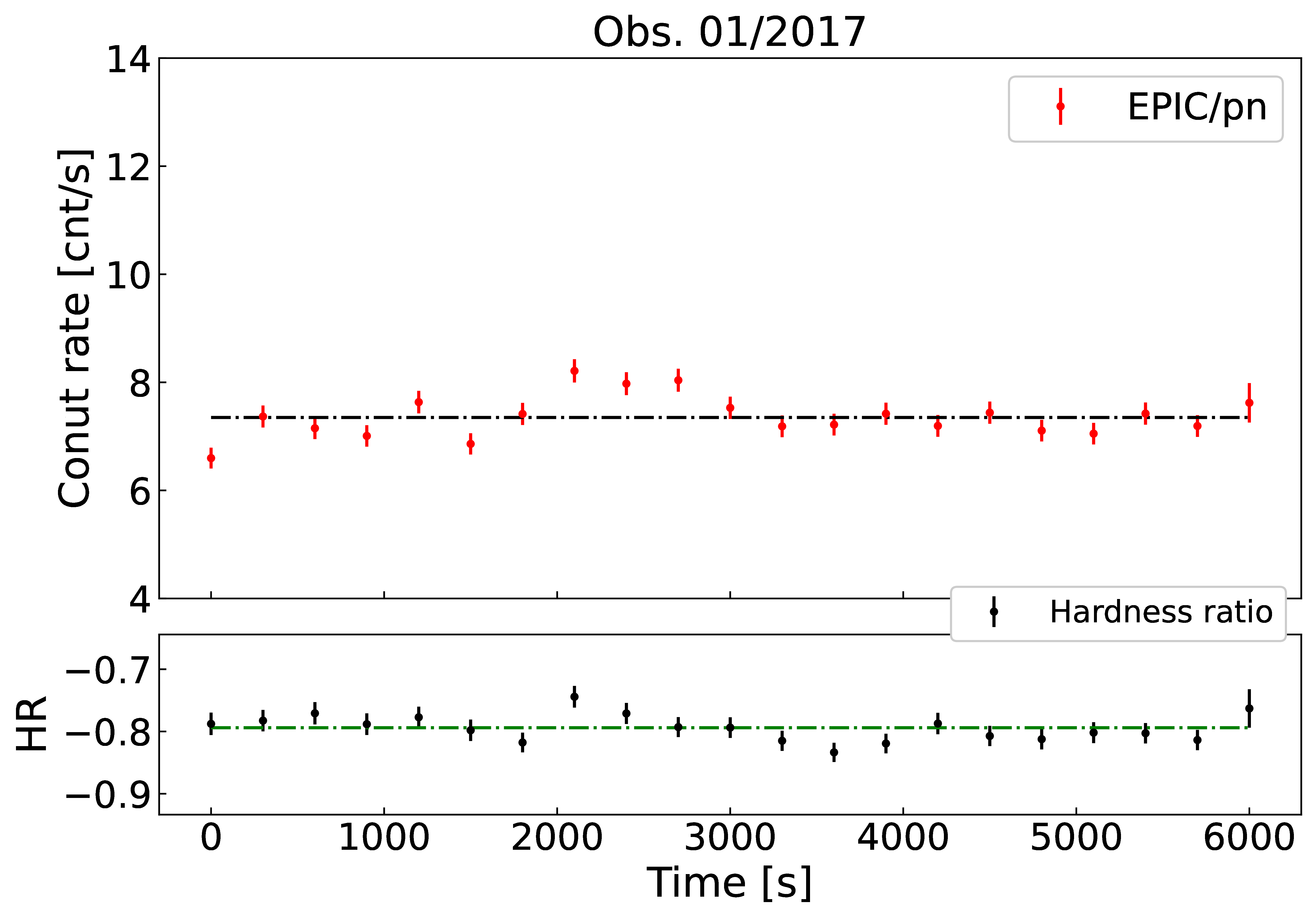}\\
\includegraphics[width=0.5\textwidth,clip]{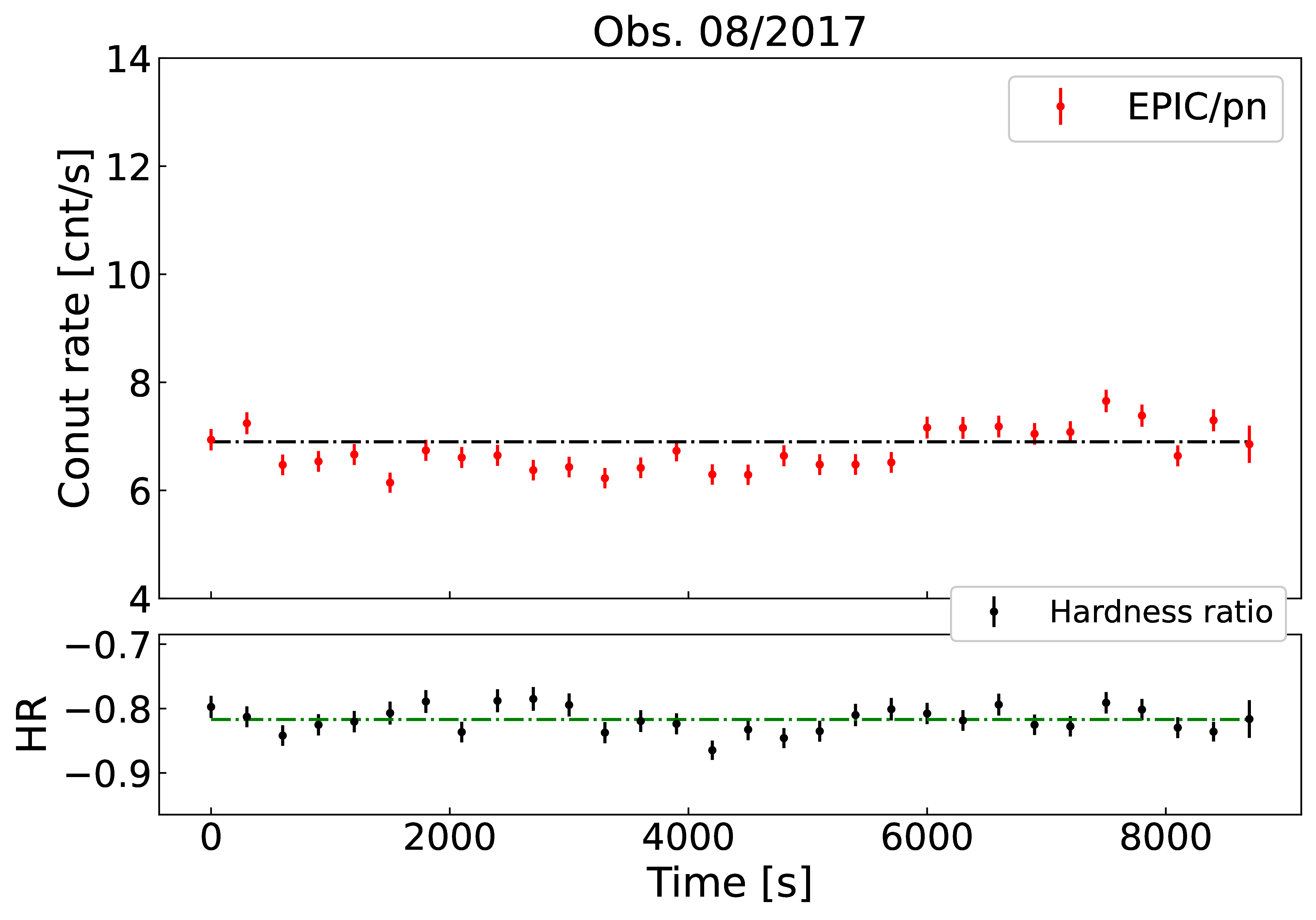}
\includegraphics[width=0.5\textwidth,clip]{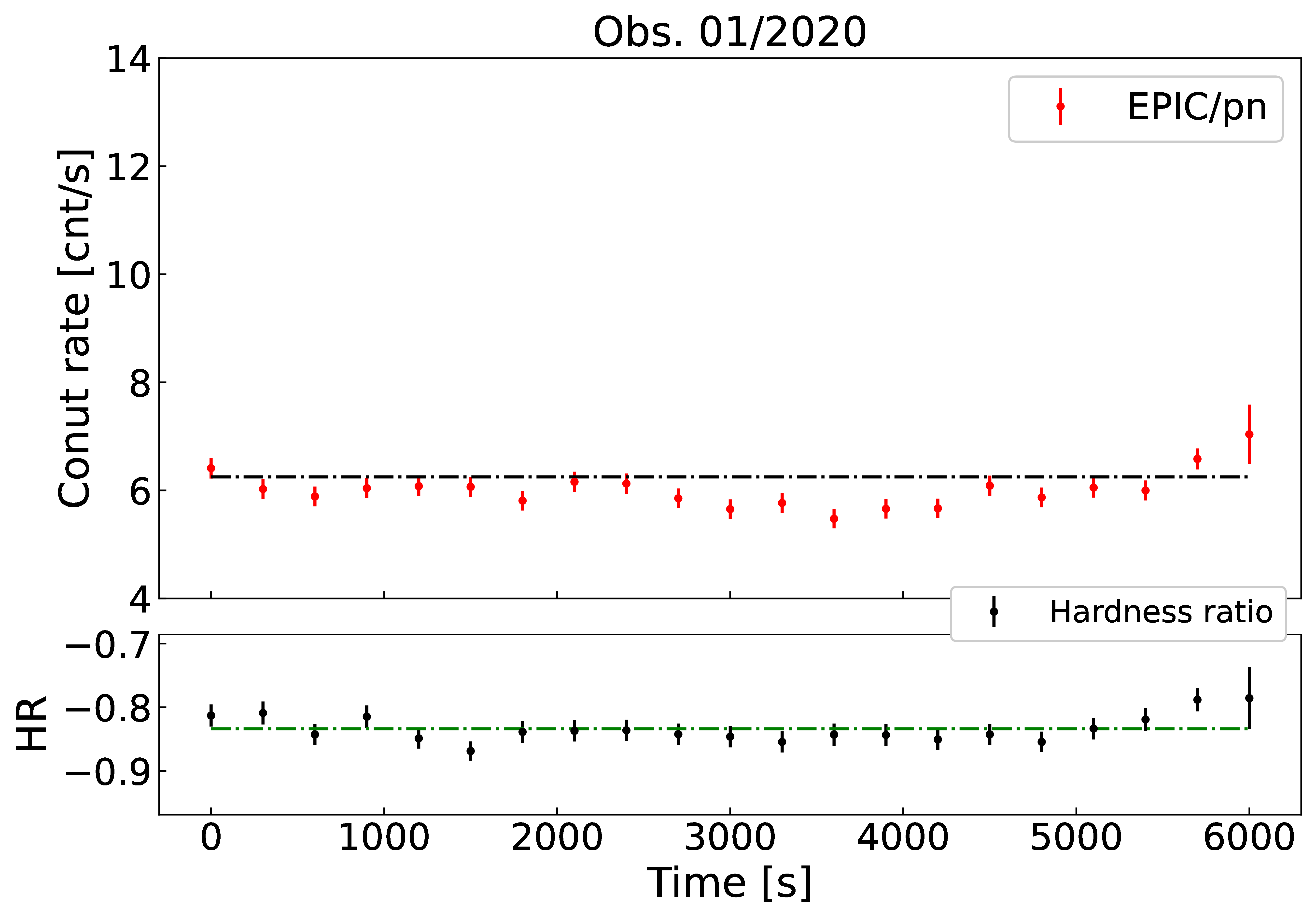}\\
\includegraphics[width=0.5\textwidth,clip]{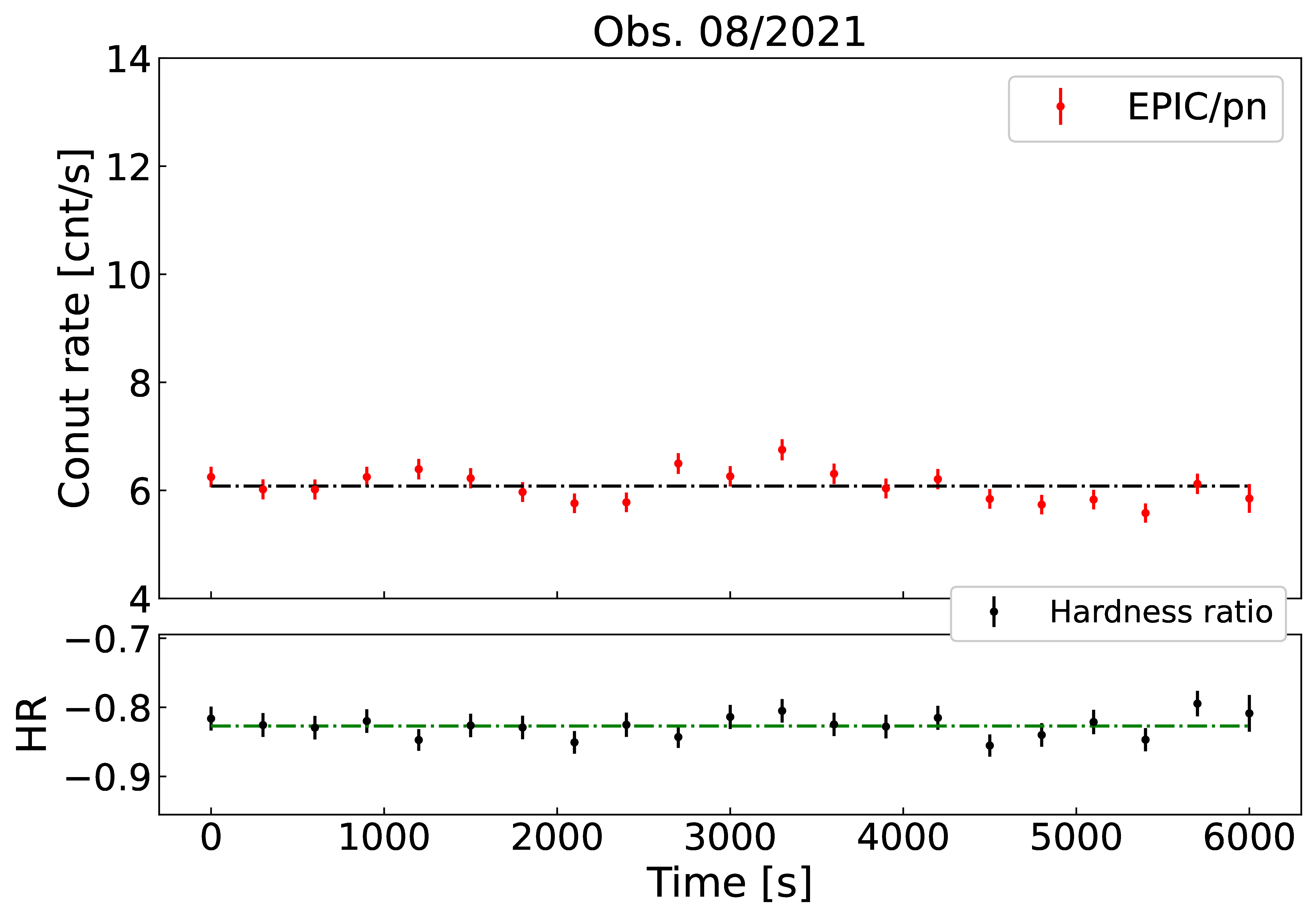}
\caption{\label{fig:hr_lc_w} Same as in Fig. \ref{fig:hr_lc}, but for the light curves without variability. 
 }
\end{center}
\end{figure*}

We present the light curves of the X-ray observations with short-term variability
that were detected with
the software \texttt{R} package {\sc changepoint} (see Sect. 3.1) in Fig. \ref{fig:hr_lc}. The light curves without short-term variability are shown in Fig. \ref{fig:hr_lc_w}. The bottom panel in each plot shows the  light curve of the hardness ratio, which was calculated from the $0.2-1.0$\,keV soft band and the $1.0-2.0$\,keV hard band, as defined in Eq.~\ref{eq:hr}. The horizontal dash-dotted lines in each panel denote the 
time segments of the constant count rate
that was identified with the {\sc changepoint} analysis.

\begin{figure*}[!htbp]
\begin{minipage}{\textwidth}
\centering
\subfloat{\includegraphics[width=0.4\textwidth]{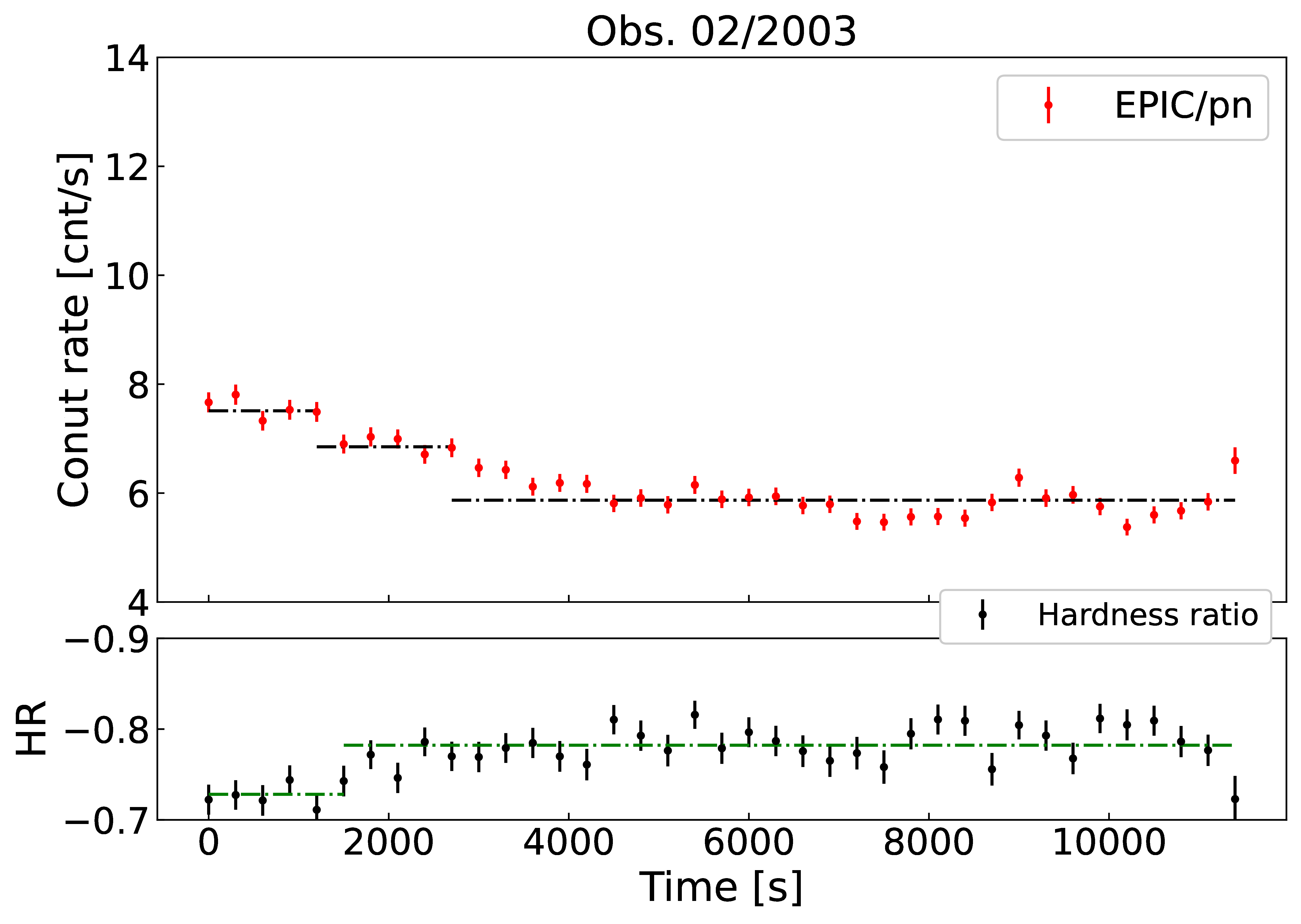}}
\subfloat{\includegraphics[width=0.4\textwidth]{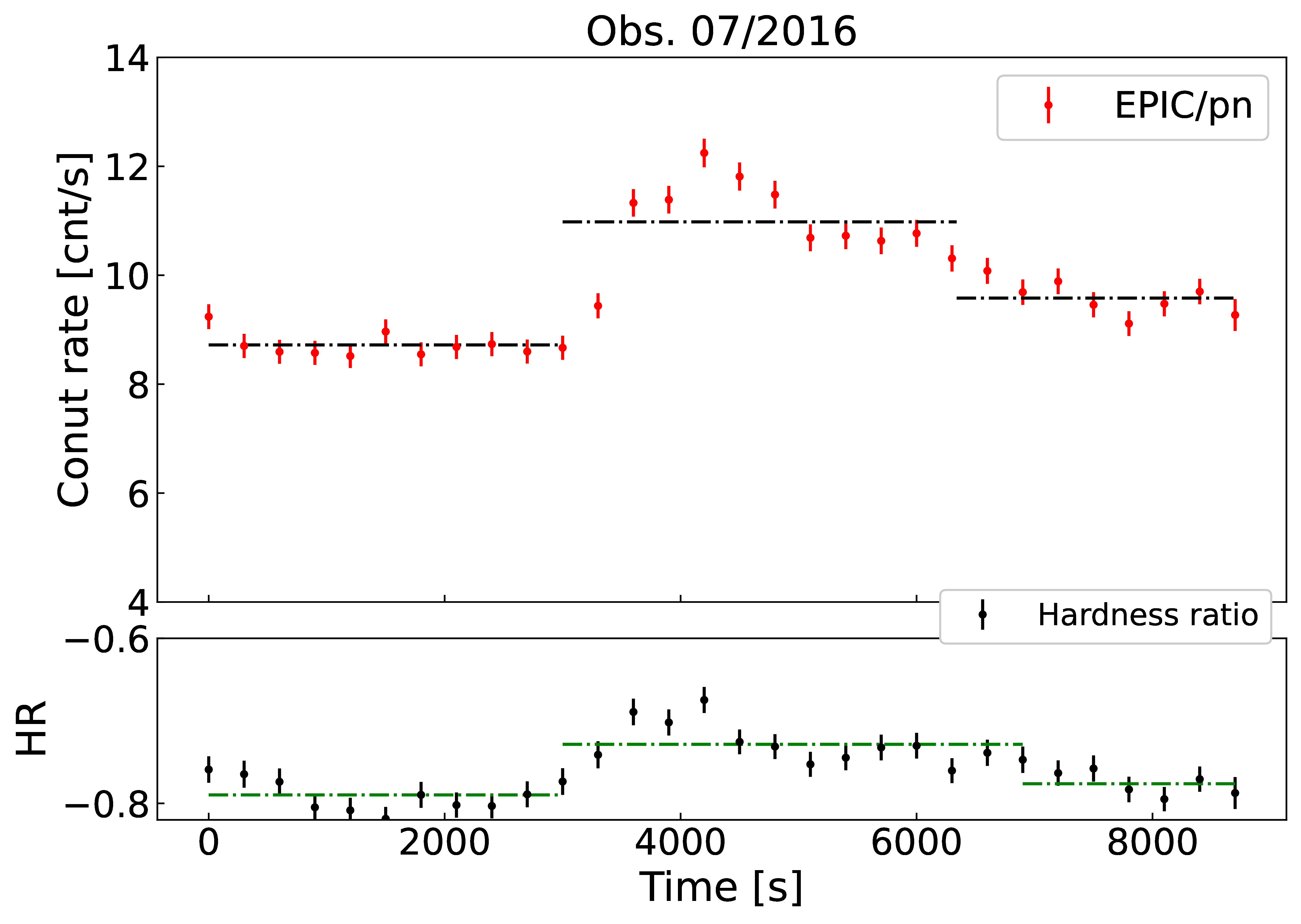}}\\
\subfloat{\includegraphics[width=0.4\textwidth]{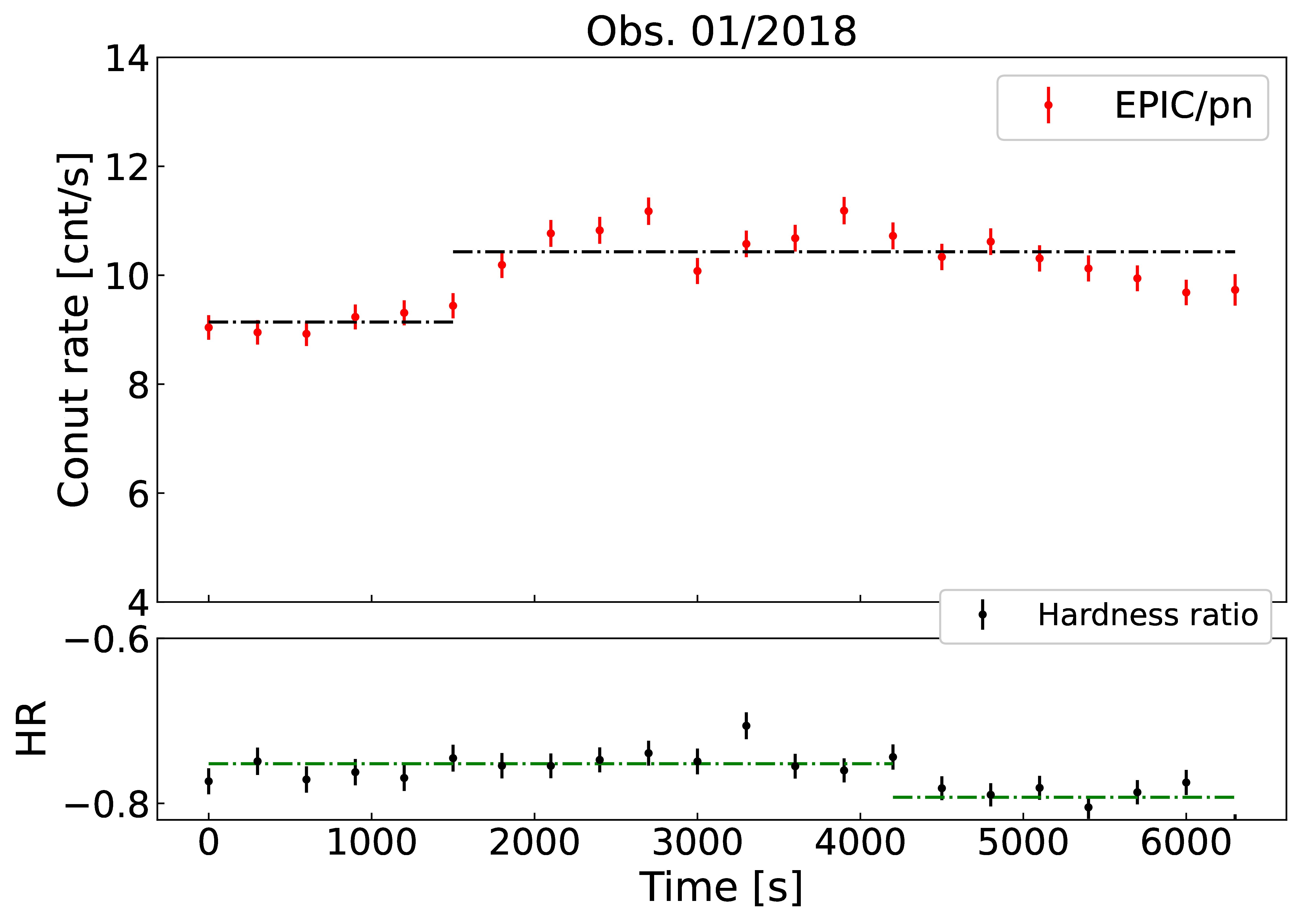}}
\subfloat{\includegraphics[width=0.4\textwidth]{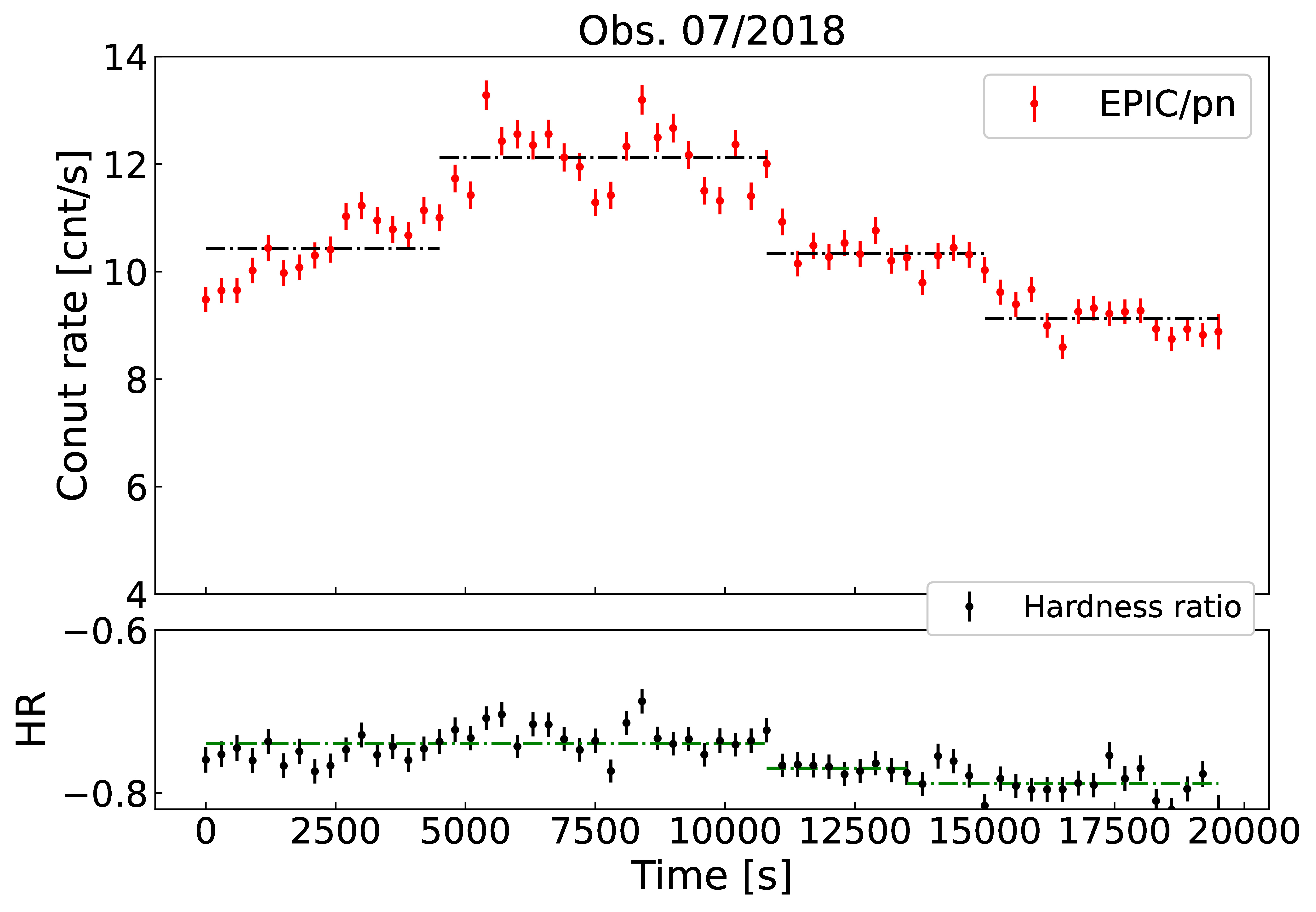}}\\
\subfloat{\includegraphics[width=0.4\textwidth]{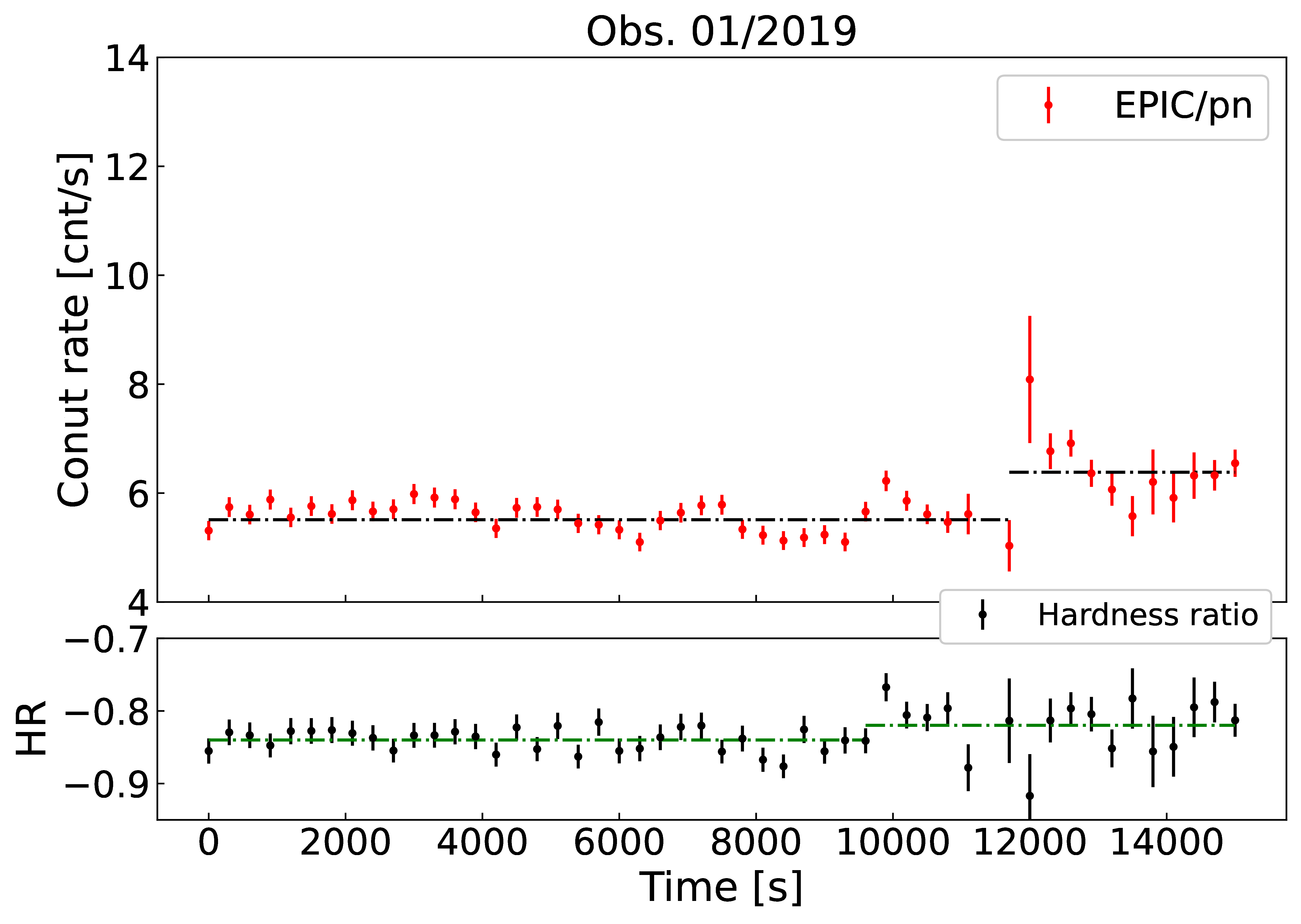}}
\subfloat{\includegraphics[width=0.4\textwidth]{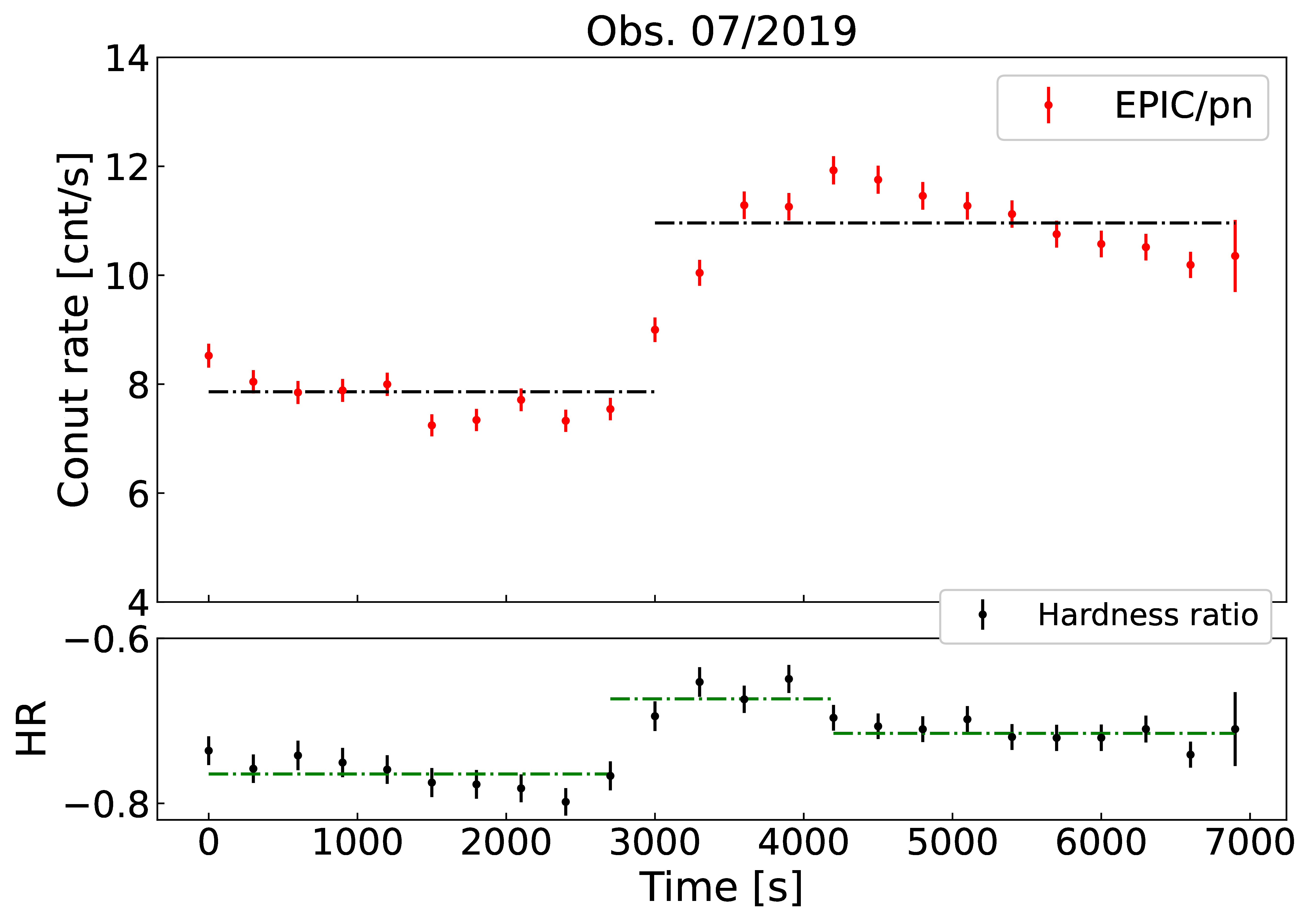}}\\
\subfloat{\includegraphics[width=0.4\textwidth]{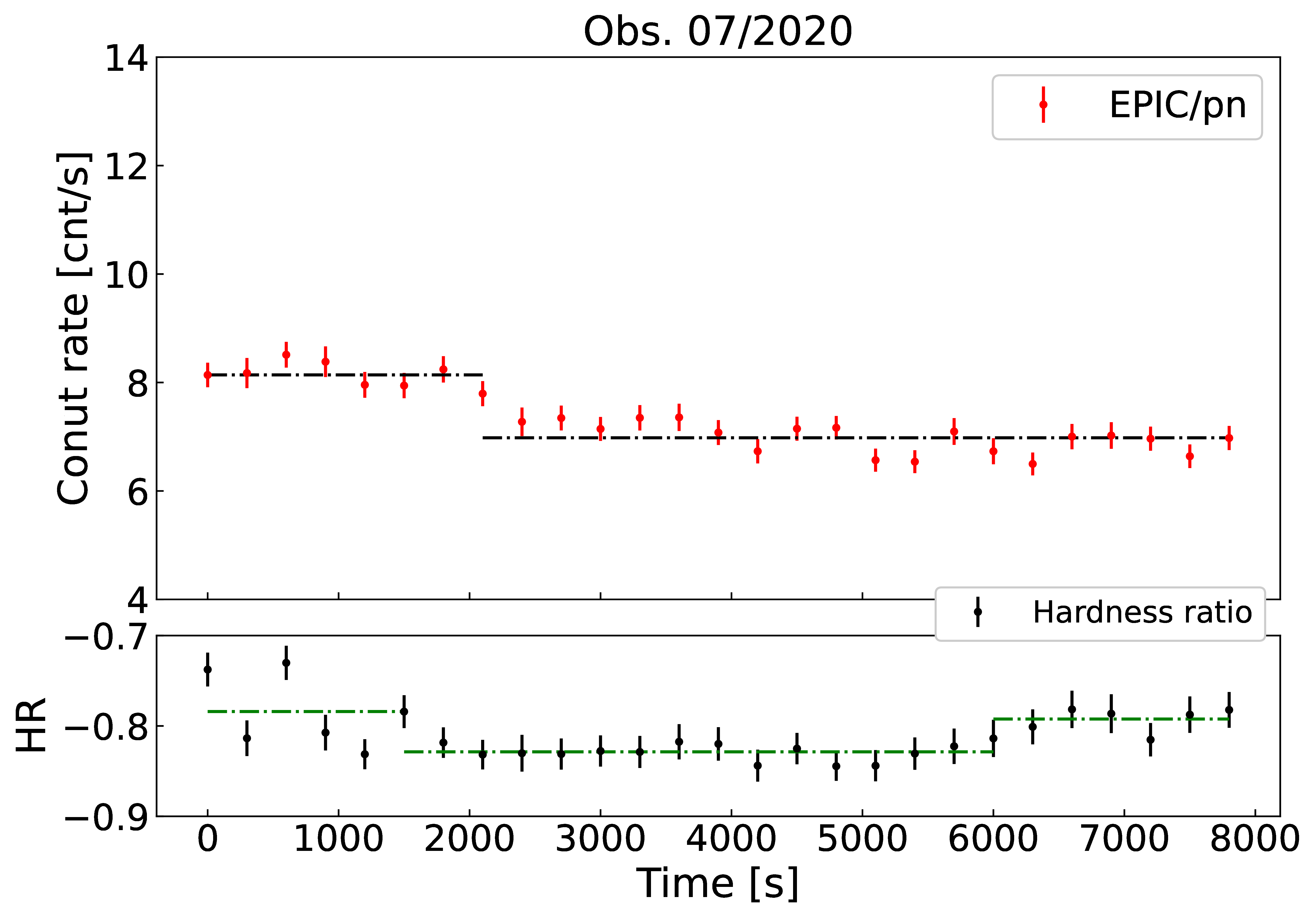}}
\subfloat{\includegraphics[width=0.4\textwidth]{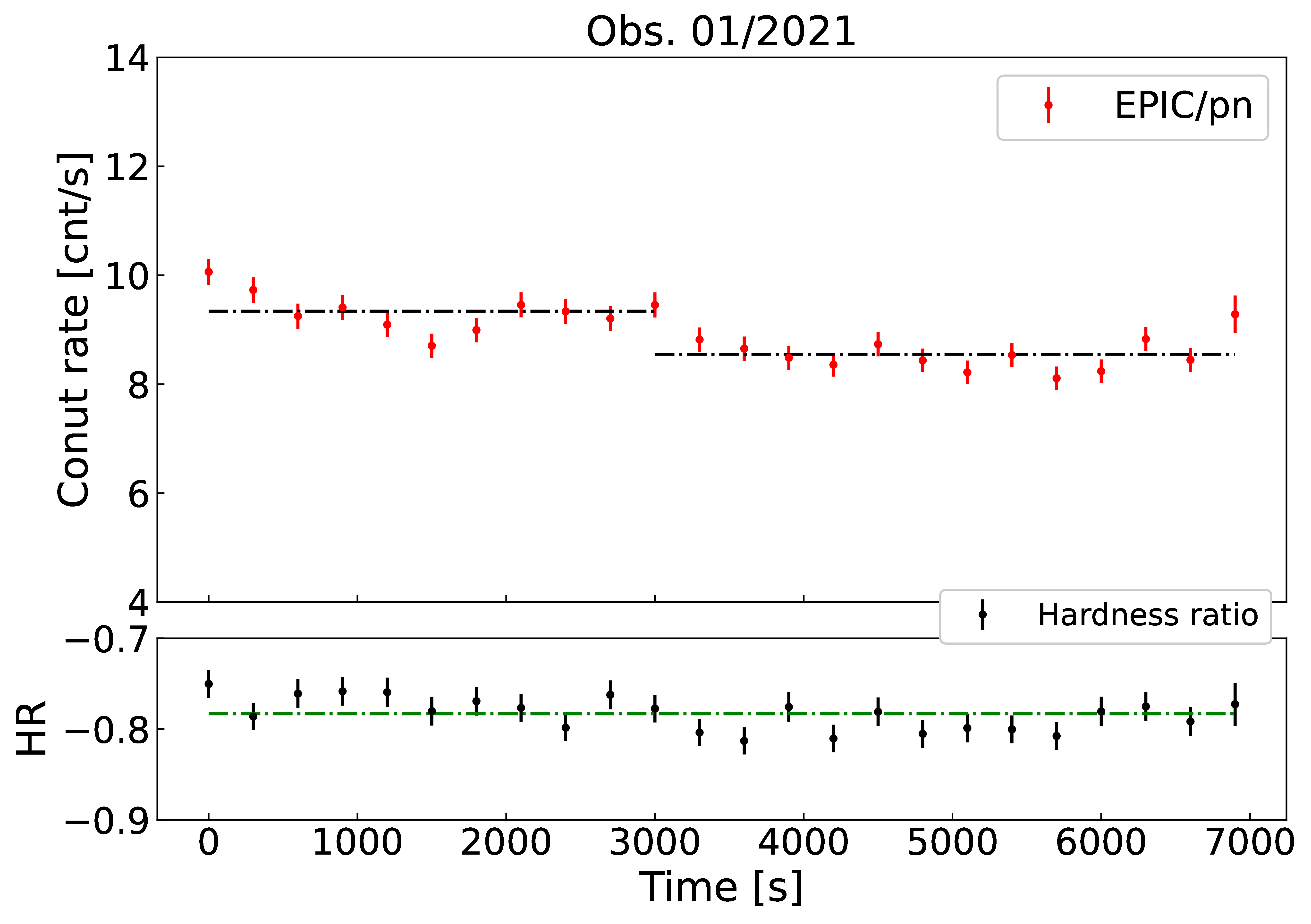}}\\
\phantomcaption 
\label{fig:hr_lc}
\end{minipage}
\end{figure*}

\begin{figure*}[!htbp]
\ContinuedFloat
\centering
\subfloat{\includegraphics[width=0.4\textwidth]{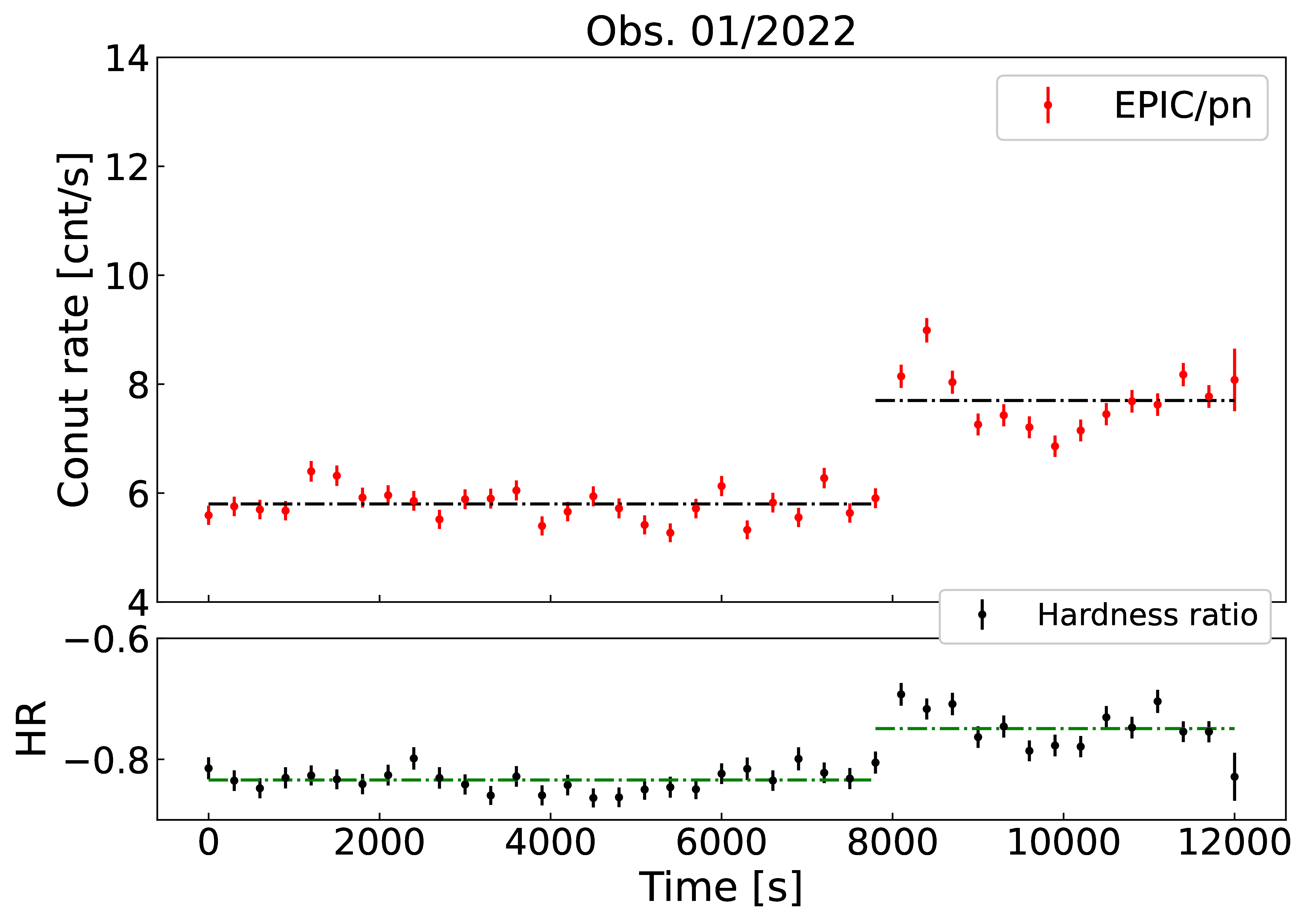}}\,
\caption{Light curves of \epsEri\, extracted in the $0.2 - 2.0$\,keV energy band
and binned with a bin size of $300$\,s. All observations 
that are flagged
as variable are shown. The individual constant segments are marked with a dash-dotted horizontal black line.
For each observation, the respective hardness ratio and its variation are shown in the lower panels. 
}
\label{fig:hr_lc}
\end{figure*}

\end{appendix}

\end{document}